\documentclass[letterpaper, aps, tightenlines, nofootinbib, 11pt]{article}
\pdfoutput = 1

\usepackage{setspace}
\usepackage[bookmarks = false, colorlinks = true, linkcolor = blue, citecolor = purple]{hyperref}
\usepackage{shortcuts}
\usepackage{titlesec}
\usepackage{cite}
\usepackage{tocloft}
\usepackage[vcentermath]{youngtab}
\usepackage{tensor}

\usepackage{tikz}
\usetikzlibrary{arrows,decorations.pathmorphing,backgrounds,positioning,fit,petri}

\usepackage[margin = 2.5cm]{geometry}
\begin{document}

\begin{center}

\thispagestyle{empty}

\begin{flushright}
\texttt{SU-ITP-15/13}
\end{flushright}

\vspace*{5em}

{\LARGE \bf Entanglement in Weakly Coupled\\ Lattice Gauge Theories}

\vspace{1cm}

{\large \DJ or\dj e Radi\v cevi\'c}
\vspace{1em}

{\it Stanford Institute for Theoretical Physics and Department of Physics\\ Stanford University \\
Stanford, CA 94305-4060, USA}\\
\vspace{1em}
\texttt{djordje@stanford.edu}\\

\vspace{0.08\textheight}
\begin{abstract}
We present a direct lattice gauge theory computation that, without using dualities, demonstrates that the entanglement entropy of Yang-Mills theories with arbitrary gauge group $G$ contains a generic logarithmic term at sufficiently weak coupling $e$. In two spatial dimensions, for a region of linear size $r$, this term equals $\frac12 \dim(G) \log(e^2 r)$ and it dominates the universal part of the entanglement entropy. Such logarithmic terms arise from the entanglement of the softest mode in the entangling region with the environment. For Maxwell theory in two spatial dimensions, our results agree with those obtained by dualizing to a compact scalar with spontaneous symmetry breaking.
\end{abstract}
\end{center}

\newpage

\tableofcontents

\section{Introduction}

Entanglement entropy is a powerful tool for characterizing the entanglement structure of quantum states. It is a quantity of interest in high energy physics, condensed matter, and quantum information theory alike. Despite its relevance, the definition of entanglement entropy in gauge theories has not been understood in depth until rather recently. The ubiquitous difficulty lay in the fact that gauge invariance introduced a degree of nonlocality at the UV scale that seemed to make it impossible to define subsystems whose entanglement entropy we were to measure. A number of approaches have been proposed to address this issue \cite{Balachandran:2013, Ohya:2004, Aoki:2015bsa, Casini:2013rba, Radicevic:2014kqa, Ghosh:2015iwa, Casini:2014aia, Levin:2006zz, Buividovich:2008yv, Donnelly:2011hn, Donnelly:2012st, Donnelly:2014gva, Buividovich:2008gq, Klebanov:2011td, Solodukhin:2011gn, Nishioka:2009un, Agon:2013iva, Gromov:2014kia, Chen:2015kfa, Hamma:2005zz, Kitaev:2005dm, Velytsky:2008rs, Yao:2010, Swingle:2012}; in subsequent sections we will review them and see how they all revolve around constructing a gauge-invariant density operator whose von Neumann entropy can be interpreted as the entanglement entropy of the given subsystem. This paper will take the route very close to the one outlined in \cite{Aoki:2015bsa, Casini:2013rba, Radicevic:2014kqa, Ghosh:2015iwa, Casini:2014aia}.

The main goal of this paper is to use these technical developments to further our understanding of the ground state entanglement in weakly coupled Yang-Mills theories. Working within a lattice gauge theory framework, we provide a comprehensive description of the calculation of entanglement entropy for Yang-Mills theories on arbitrary lattices and for arbitrary gauge groups $G$. In particular, we demonstrate the presence of a ubiquitous term that, in a continuum with $d$ spatial dimensions, takes the form
\bel{
  \Delta S = \frac12 (d - 1)\, \dim(G) \, \log \left(e^{\frac2{3 - d}} r\right).
}
Here $r$ is the linear size of the entanglement region and $e$ is the continuum gauge coupling, and for $d = 3$ the logarithm is replaced just by $\log e^2$. This term is of particular importance in $d = 2$, where it is the dominant universal term of the entanglement entropy that takes the form
\bel{\label{eq Delta S ad}
  \Delta S = \frac12 \dim(G)\, \log\left(e^2 r\right).
}

We obtain the advertised results by gauge-fixing to axial gauge, expressing the gauge theory as a principal chiral model with spontaneous symmetry breaking, and calculating the entanglement entropy of the resulting Nambu-Goldstone bosons.\footnote{A clarification is in order here.  Typically, a discussion of spontaneous symmetry breaking in an $O(N)$ model of scalar matter assumes that the constant mode is frozen into a particular position on the sphere $S^{N - 1}$. However, one can also construct the state that features ``restoration of symmetry'' by uniformly superposing all possible directions that the constant mode can point in. This construction effectively ``gauges away'' the zero mode of the theory, and it is this kind of state that we will encounter when gauge-fixing the Yang-Mills theory. In general, whenever we refer to ``spontaneous symmetry breaking'' in this paper, we will refer to a projection to just the sector of Nambu-Goldstone bosons, where the wavefunction does not depend on the zero mode.} The key ingredient here is the fact that Nambu-Goldstone bosons do not have a zero mode and therefore exhibit enhanced entanglement of the softest mode in the entanglement region \cite{Metlitski:2011}. Thus, we can qualitatively say that the $\Delta S$ term arises because weakly coupled gauge theories look like $(d - 1) \dim(G)$ decoupled photons, with each photon behaving like a scalar field with a zero mode removed (or gauged away). For the special case of the $d = 2$ Maxwell theory, this logarithmic term has been computed in an alternative way, by  dualizing to a compact scalar theory (the $O(2)$ model) that exhibits spontaneous symmetry breaking \cite{Agon:2013iva}.\footnote{As we will describe in detail below, the gauge-fixed lattice theory is related to the dual scalar one by a canonical transformation.}

The results presented in this paper touch on many other studies of entanglement in gauge theories. Our computation shows that the $\Delta S$ term is at least one part of entanglement entropy that is invariant under field-theoretic dualities. Further, the weak-coupling entanglement entropy is explicitly shown to scale as $N^2$ in the planar limit, forming a contrast to the vanishing of the entropy at strong lattice coupling and suggesting, along the lines of \cite{Klebanov:2007ws}, that entanglement entropy is indeed a good order parameter for confinement that can be explicitly computed in both weak and strong coupling regimes. Finally, we draw a connection between the logarithmic terms above and the topological entanglement entropy in $d = 2$ \cite{Kitaev:2005dm, Levin:2006zz}, arguing that $\Delta S$ is the analog of this important quantity for systems with continuous gauge groups.

The paper is structured as follows. In Section \ref{sec notation}, we give a short but self-contained introduction to lattice gauge theories and set the notation to be used throughout the paper. In Section \ref{sec prelim}, we give a very explicit definition of entanglement entropy in a gauge theory \cite{Aoki:2015bsa, Casini:2013rba, Radicevic:2014kqa, Ghosh:2015iwa, Casini:2014aia}, and we show how it agrees with other definitions in the literature. An example is given in Section \ref{sec strong}, where the strong coupling entanglement entropy at large $N$ is computed for the first time. The main calculation is given in Section \ref{sec weak} and it culminates with the above result for $\Delta S$ and with a reasonable conjecture for the general form of the total gauge theory entropy. Further implications of our results are discussed in the Conclusion, and miscellaneous technical points are collected in the Appendices.

\section{Notation and conventions} \label{sec notation}

Our Hamiltonian formulation of non-Abelian lattice gauge theory in $d$ spatial dimensions is based on the seminal work of Kogut and Susskind \cite{Kogut:1974ag}. The notation follows the approaches of \cite{Radicevic:2014kqa, Casini:2013rba}. We work on a finite lattice with open boundary conditions and no nontrivial topology.  The lattice sites are labeled by $i, j, \ldots$, and the links are labeled either by a link index, $\ell$, by a pair of adjacent site indices, $(i,j)$, or by a site and a direction, $(i,\mu)$. Each link is oriented, so if $\ell = (i,j)$, the link of opposite orientation is ${\bar\ell} = (j, i)$. In all examples we will assume a hypercubic lattice, but our discussion applies to arbitrary lattices.

Quantum variables live on links. The state on a link $\ell$ is labeled by an element $U$ of the gauge group $G$. Products of states $\qvec U_\ell$ over all links $\ell$ form the Hilbert space $\H_0$ of the whole lattice. The operator algebra on $\H_0$ is generated by \textbf{momentum operators} $L^\Lambda_\ell$ and \textbf{position operators} $U^r_\ell$, which act on $\qvec U_\ell$ via
\bel{\label{pos mom ops}
   L^\Lambda_\ell \qvec U_\ell = \qvec {\Lambda U}_\ell,\quad  U^r_\ell \qvec U_\ell = r(U) \qvec U_\ell,
}
where $\Lambda \in G$ and $r$ is a representation of the gauge group.\footnote{There exists a second set of momentum operators in nonabelian theories --- it acts by right-multiplication by $\Lambda$.  Its existence will not be used in this paper.} We also define $L^\Lambda_{{\bar\ell}}$ and $ U^r_{{\bar\ell}}$ via
\bel{
   L^\Lambda_{{\bar\ell}} \qvec U_\ell = \qvec{U \Lambda^{-1}}_\ell, \quad  U^r_{{\bar\ell}} \qvec U_\ell = r(U^{-1}) \qvec U_\ell.
  }
Note that $L^\Lambda_\ell$ and $L^\Lambda_{\bar\ell}$ commute, by construction. Operators on different links also all commute. By writing $L_\ell$  we also refer to the operator $\left( \prod_{\ell' \neq \ell} \1_{\ell'}\right) \times L_\ell$ on the full lattice. The state of the entire lattice (i.e.~an element of $\H_0$) will be denoted by kets without an index; e.g.~we might write $\qvec \Psi = \prod_\ell \qvec{U}_\ell$ for a given configuration $\{U_\ell\}$ on the lattice links.

\textbf{Electric operators} $J^a_\ell$ are representations of generators associated to momentum operators:
\bel{\label{def J}
  L^\Lambda_\ell \equiv e^{i\theta^a J_\ell^a} \quad \trm{for} \quad \Lambda \equiv e^{i\theta^a T^a}.
}
Here $\theta^a$ are the coordinates on the group manifold and $T^a$ are the generators of $G$ normalized to $\Tr(T^a T^b) = \delta^{ab}$.\footnote{The summation convention applies to indices $a$, $b$, etc, but \emph{not} to lattice indices $\ell$, $i$, $j$, $p$, etc.} Given a quantum state labeled by $U \equiv e^{iA^aT^a}$, the electric operators act on it as covariant derivatives on the Lie manifold with coordinates $A^a$:
\bel{
  J_\ell^a \qvec U_\ell = \frac{L^\eps_\ell - \1}{i\eps^a}\qvec{e^{iA^a T^a}}_\ell = \frac1{i\eps^a} \left(\qvec{e^{i\left(A^a + \eps^a + \frac i2 f^{abc}\eps^b A^c\right)T^a}}_\ell - \qvec{e^{iA^a T^a}}_\ell \right) \equiv \frac1i \frac{\trm D}{\trm DA^a} \qvec{U}_\ell.
}
The covariant Laplacian $\b J^2_\ell =  J^a_\ell J^a_\ell$ is a Casimir invariant that we will use extensively. The sum $\sum_\ell \b J^2_\ell$ is the ``electric term'' in the Kogut-Susskind Hamiltonian \cite{Kogut:1974ag}. The remaining, ``magnetic term'' of this Hamiltonian is given by the \textbf{magnetic operators} $W_p^r$ and $\overline W_p^r$, defined on lattice plaquettes $p$ as
\bel{
  W^r_p \equiv \Tr\bigg(\prod_{\ell \in p} U^r_\ell \bigg) \quad \trm{and} \quad \overline W^r_p \equiv \Tr\bigg(\prod_{\ell \in p} U^r_{\bar\ell} \bigg),
}
where $r$ is a representation of the gauge group and the links in the product $\prod_{\ell \in p}$ are traversed counterclockwise.

Henceforth, when $r$ is dropped, the fundamental representation is understood. We will always use $N$ to denote the dimension of this representation. Using these conventions, the full Kogut-Susskind Hamiltonian is
\bel{\label{H}
  H = g^2 \sum_\ell \b J_\ell^2 + \frac1{g^2}\sum_p \left[2N - \left(W_p + \overline W_p\right)\right],
}
where $g^2$ is the gauge coupling. We will exclusively work with this Hamiltonian in this paper, though our results would not qualitatively change if we included more terms in the magnetic potential.

Gauge-invariant systems are insensitive to all local transformations of the form
\bel{
  U_{(i,\,j)} \mapsto U_{(i,\,j)}^{\boldsymbol{\Lambda}} \equiv \Lambda_i U_{(i,\,j)} \Lambda_j^{-1},
}
where $\boldsymbol{\Lambda} = \{\Lambda_i\}$ assigns an element of $G$ to each lattice site. Such transformations are implemented in terms of operators via
\bel{\label{def G}
  \qvec{\Psi^{\boldsymbol{\Lambda}}} \equiv G^{\boldsymbol{\Lambda}} \qvec \Psi,\quad G^{\boldsymbol{\Lambda}} \equiv \prod_i G_i^{\Lambda_i} \equiv \prod_{i,\, \mu} L_{(i,\,\mu)}^{\Lambda_i},
}
where we have introduced the \textbf{Gauss operators} $G_i^\Lambda \equiv \prod_\mu L^\Lambda_{(i,\, \mu)}$ at each site. In the above product there are exactly two momentum operators acting on each link, one acting in the direction of the link and one in the opposite direction. Together they implement the desired local transformation. Gauging this transformation, we demand that physical states are only those satisfying $\qvec{\Psi^{\boldsymbol{\Lambda}}} = \qvec \Psi$ or, equivalently, $G_i^\Lambda\qvec \Psi = \qvec \Psi$ at each site and for any $\Lambda \in G$. The space of all such states is the physical Hilbert space $\H$.

In the first half of the paper we will work in the \textbf{electric basis} of $\H_0$ and $\H$. This basis diagonalizes the electric term $\b J^2_\ell$ on each link. One element of this basis is the ground state of the electric term,
\bel{\label{def Omega}
  \qvec \Omega = \prod_\ell \qvec\Omega_\ell, \quad \qvec \Omega_\ell \equiv \int_G \d U \qvec U_\ell,
}
where $\d U$ is the Haar measure on the group manifold $G$, normalized so that $\qprod\Omega\Omega = 1$.\footnote{This normalization implies that $\int_G \d U = \sqrt{\trm{Vol}(G)}$ and $\int_G \d U\, \delta(U - V) = 1/{\sqrt{\trm{Vol}(G)}}$.} This state is gauge-invariant, as $L_\ell^\Lambda\qvec\Omega = \qvec\Omega$ for all momentum operators $L_\ell^\Lambda$. Excited states --- other elements of the electric basis --- are formed by acting on $\qvec \Omega$ with position operators $U_\ell^r$. These excitations are labeled by representations $r$, as per the Peter-Weyl theorem. Physical excitations can be viewed as closed lines of electric (color) flux. In the remainder of this subsection we review the systematics of these electric eigenstates. An excellent reference with many more details is \cite{Drouffe:1983fv}.

Elements of the electric basis of $\H_0$ can be thought of as products over links of wavefunctions on the space of representations of $G$. On a given link, an electric basis element labeled by an irreducible representation $r$ is
\bel{
  \qvec{r}_\ell \equiv \sqrt{d_r} \int_G \d U\, U^r \qvec U_\ell,
}
where $d_r$ is the dimension of $r$. The normalization $\tensor[_\ell]{\qprod{r}{r'}}{_\ell} = \delta_{rr'}\, r(\1)$ follows from eq.~\eqref{def Omega} and the Weyl orthogonality property
\bel{
  \int_G \d U\, \big(U^{r'}\big)^* \otimes U^r = \frac{\sqrt{\trm{Vol}(G)}}{d_r} \, \delta_{rr'}\, r(\1).
}
As an example, the fundamental and antifundamental representations $r = \trm f,\, \bar{\trm f}$ are
\bel{
  \qvec{\trm f_{\alpha\beta}}_\ell \equiv \sqrt{N} \int_G \d U\, U_{\alpha\beta} \qvec U_\ell,\quad   \qvec{\bar{\trm f}_{\alpha\beta}}_\ell \equiv \sqrt{N} \int_G \d U\, \big(U^{-1}\big)_{\beta\alpha} \qvec U_\ell
}
with $\alpha, \beta = 1, 2,\ldots N$.

Given a (possibly self-intersecting or multiply-winding) closed loop of links $\mathcal C = (\ell_1, \ldots, \ell_n)$, the physical states on $\mathcal C$ are
\bel{\label{def square}
  \qvec{\square^r_{\mathcal C}} = \int_G \d U_1\ldots \d U_n\ \Tr_r \left(U_1^{s_1} \ldots U_n^{s_n}\right) \ \qvec{U_1}_{\ell_1} \ldots \qvec{U_n}_{\ell_n} = \frac1{d_r^{n/2}} \Tr\big(\qvec{r^{s_1}}_1 \ldots \qvec{r^{s_n}}_n \big)
}
where $s_k = + 1$ if the link $\ell_k \in \mathcal C$ is traversed in the direction of its orientation, and $s_k = -1$ if not. These states form the electric basis of $\H$; they are eigenstates of the electric term of \eqref{H} with eigenvalues $n g^2 C_2(r)$, where $C_2(r)$ is the quadratic Casimir of the representation $r$ of the gauge algebra. For example, we have $C_2(\trm f) = C_2(\bar{\trm f}) = N$ for $U(N)$ and $C_2(\trm f) = C_2(\bar{\trm f})= N - 1/N$ for $SU(N)$. At large $N$, states defined on different loops are all orthogonal to each other, and hence electric basis elements in the planar limit are indexed by the set of all closed loops with a representation associated to each loop.

\section{Entanglement entropy in lattice gauge theory} \label{sec prelim}

\subsection{Overview} \label{subsec overview}

Entanglement entropy quantifies how much information is lost by restricting ourselves to a part of the given system. The point of view of this paper is that, given a state of the whole system, one can always construct a reduced density operator that precisely reproduces the physics (i.e.~all the correlation functions) of the original state in just a part of the full system. The entanglement entropy is then the von Neumann entropy associated to this operator; roughly speaking, this measures the number of states a subsystem can be in while the entire system is in a given state. This is the quantity we will compute. There exist alternative (possibly inequivalent) approaches to entanglement entropy, e.g.~axiomatic definitions involving strong subadditivity, but we will not address them here.

The definition of a reduced density operator in gauge theories is tricky because the physical Hilbert space $\H$ does not admit a decomposition into a direct product of Hilbert spaces defined solely on a region $V$ and its complement $\bar V$. There are several ways to address this subtlety:
 \begin{itemize}
   \item Embedding $\H$ into a direct product of Hilbert spaces defined separately on $V$ and $\bar V$ gives rise to a density operator that can be reduced in the usual way \cite{Levin:2006zz, Buividovich:2008yv, Donnelly:2011hn, Donnelly:2012st, Donnelly:2014gva, Buividovich:2008gq}.
   \item Elements of the reduced density matrix may be computable via a Euclidean path integral by directly path-integrating and using the replica trick (see e.g.~\cite{Klebanov:2011td, Agon:2013iva, Gromov:2014kia} for some salient examples).
   \item In theories that admit holographic duals, entanglement entropy can be computed following the Ryu-Takayanagi prescription \cite{Solodukhin:2011gn, Nishioka:2009un} (see \cite{Lewkowycz:2013nqa} for an explanation on why this is equivalent to the computation via the replica method in the boundary theory).
   \item In special circumstances, the gauge theory setup can be mapped to a problem in which the reduced density operator/entanglement entropy calculation is tractable \cite{Chen:2015kfa, Hamma:2005zz, Kitaev:2005dm, Velytsky:2008rs, Yao:2010, Swingle:2012}. These approaches might not yield the most general method for computing a reduced density matrix, but they are important for checking any general prescription.
   \item Finally, a reduced density operator can be defined purely algebraically, as the unique density operator in any subalgebra of observables that reproduces the expectation values of all the operators in the subalgebra \cite{Aoki:2015bsa, Ghosh:2015iwa, Radicevic:2014kqa, Casini:2014aia, Casini:2013rba, Balachandran:2013, Ohya:2004}.
 \end{itemize}

These methods are equivalent in a precise sense that we will discuss below. We will start from the particularly transparent approach taken in the progression of papers \cite{Ghosh:2015iwa, Radicevic:2014kqa, Aoki:2015bsa, Casini:2013rba}, where the last of the above prescriptions was followed. Given the algebra $\A_V$ of all gauge-invariant operators on a set of links (not sites or plaquettes) $V$, it is \emph{always} possible to find the density operator $\rho_V \in \A_V$ and calculate its von Neumann entropy. This prescription is manifestly gauge-invariant and associates an entropy to all the physical data contained in a region of space\footnote{The issue of taking a continuum limit of a set of links may sound delicate; we will comment on this below.}. Varying what algebra one assigns to a region (e.g.~dropping magnetic operators near the edges of the set $V$) leads to different values for the entanglement entropy \cite{Casini:2013rba, Casini:2014aia}; the results of \cite{Ohmori:2014eia} suggest that these alternative entropies correspond to entanglement regions with operator insertions on the entangling edge, and we will have more to say about this at the end of this Section.

\subsection{Definition}

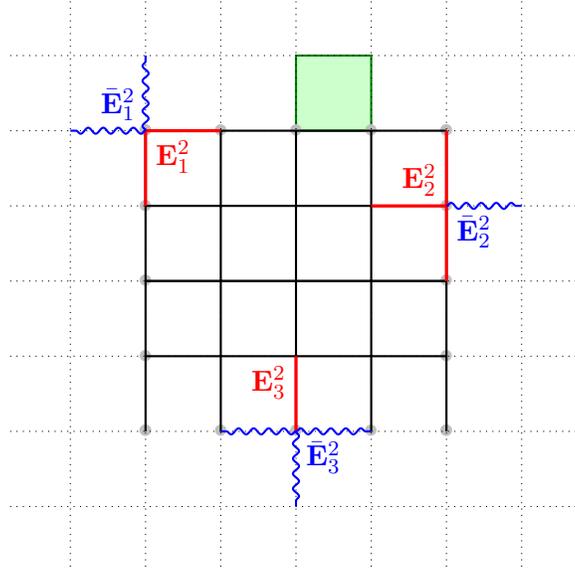
\begin{figure}[tb!]
\begin{center}

\begin{tikzpicture}[scale = 2]

  \filldraw [fill=green!20, draw=green!50!black, thick] (0, 1) rectangle +(0.5, 0.5);

  \foreach \x in {-1, -0.5,..., 1} \draw (\x, -1) node[lightgray] {$\bullet$};
  \foreach \x in {-1, -0.5,..., 1} \draw (\x, 1) node[lightgray] {$\bullet$};
  \foreach \x in {-1, -0.5,..., 1} \draw (1, \x) node[lightgray] {$\bullet$};
  \foreach \x in {-1, -0.5,..., 1} \draw (-1, \x) node[lightgray] {$\bullet$};

  \draw[step = 0.5, dotted] (-1.9, -1.9) grid (1.9, 1.9);
  \draw[step = 0.5, thick] (-1, -0.5) grid (1, 1);
  \draw[xstep = 0.5, ystep = 5, thick] (-1, -1) grid (1, -0.5);

  \draw[very thick, red] (-1, 0.5) -- (-1, 1) -- (-0.5, 1);
  \draw[red] (-1, 1) node[anchor = north west] {$\b E^2_1$};
  \draw[thick, blue, decorate, decoration = {snake, amplitude = .4mm, segment length = 2mm, post length = 0.1mm}] (-1, 1) -- (-1.5, 1);
  \draw[thick, blue, decorate, decoration = {snake, amplitude = .4mm, segment length = 2mm, post length = 0.1mm}] (-1, 1) -- (-1, 1.5);
  \draw[blue] (-1, 1) node[anchor = south east] {$\bar{\b E}^2_1$};

  \draw[very thick, red] (1, 0.5) -- (0.5, 0.5);
  \draw[very thick, red] (1, 0.5) -- (1, 1);
  \draw[very thick, red] (1, 0.5) -- (1, 0);
  \draw[red] (1, 0.5) node[anchor = south east] {$\b E^2_2$};
  \draw[thick, blue, decorate, decoration = {snake, amplitude = .4mm, segment length = 2mm, post length = 0.1mm}] (1, 0.5) -- (1.5, 0.5);
  \draw[blue] (1, 0.5) node[anchor = north west] {$\bar{\b E}^2_2$};

  \draw[very thick, red] (0, -0.5) -- (0, -1);
  \draw[red] (0, -0.5) node[anchor = north east] {$\b E^2_3$};
  \draw[thick, blue, decorate, decoration = {snake, amplitude = .4mm, segment length = 2mm, post length = 0.1mm}] (0, -1) -- (0, -1.5);
  \draw[thick, blue, decorate, decoration = {snake, amplitude = .4mm, segment length = 2mm, post length = 0.1mm}] (0, -1) -- (0.5, -1);
  \draw[thick, blue, decorate, decoration = {snake, amplitude = .4mm, segment length = 2mm, post length = 0.1mm}] (0, -1) -- (-0.5, -1);
  \draw (0, -1) node[blue, anchor = north west] {$\bar{\b E}^2_3$};

\end{tikzpicture}

\end{center}
\caption{\small (\textsc{color online, adapted from \cite{Radicevic:2014kqa}}) Three examples of boundary electric operators defined in eq.~\eqref{bdry el ops}. Full lines denote links in $V$ and dotted lines denote links in $\bar V$. Gray dots denote elements of $\del V$. The boundary electric Casimirs $\b E^2_i$, defined on each boundary site, depend on electric generators on all links that enter that site and belong to $V$ (thick red lines), and conversely for $\bar {\b E}^2_i$ and the wavy blue lines. Instead of specifying values of electric boundary operators, superselection sectors can also be specified by values of magnetic boundary operators. One such operator is the Wilson loop along the green plaquette at the edge of $V$.}
\label{fig def}
\end{figure}

The desired reduced density operator is constructed as follows. Let $V$ and $\bar V$ be a set of links and its complement, and let $\del V$ be the set of sites for which some (but not all) emanating links are in $V$. For each $i \in \del V$, define \textbf{boundary electric operators} as
\bel{\label{bdry el ops}
  E_i^a = \sum_{(i,\,j) \in V} J_{(i,\,j)}^a, \quad \bar E_i^a = \sum_{(i,\,j) \in \bar V} J_{(i,\,j)}^a.
}
Gauss operators \eqref{def G} at boundary sites $i$ can be expressed as
\bel{
  G_i^\Lambda = e^{i\theta^a (E_i^a + \bar E_i^a)},
}
where, as in \eqref{def J}, the $\theta$'s are defined such that $\Lambda = e^{i\theta^a T^a}$. The gauge constraint requires $G_i^\Lambda = 1$ for all $\Lambda$, and hence it must be true that
\bel{
  E_i^a = - \bar E_i^a
}
when acting on any physical state. Of course, these operators are not gauge-invariant, but e.g.~the quadratic Casimirs $\b E_i^2 = E_i^a E_i^a$ are, and the above relation implies that
\bel{
  \b E_i^2 = \bar{\b E}_i^2.
}
All possible boundary Casimirs --- not just the quadratic ones --- generate the center of the algebra $\A_V$. The center elements are all diagonalized simultaneously in the electric basis. The physical space $\H$ thus naturally splits into superselection sectors $\H^{(\b k)}$ labeled by $\b k = (k_1, \ldots, k_B)$, the collection of Casimir eigenvalues at various sites $i$, where $B$ is the total number of boundary Casimirs. See Fig.~\ref{fig def} for an illustration.

A subtle point arises here: the space $\H^{(\b k)}$ is spanned by gauge-invariant states living wholly in $V$, wholly in $\bar V$, and partly in $V$ and partly in $\bar V$. This last group presents an obstruction to decomposing $\H^{(\b k)}$ into a direct product, and we now describe how this is circumvented by working with the example of a square plaquette $p = (\ell_1, \ell_2, \ell_3, \ell_4)$ with just $\ell_1 \in V$. Our point is illustrated already by the fundamental excitation on this plaquette, $\qvec{\square_p} = \int_G \Tr(U_1 U_2 U_3^{-1} U_4^{-1}) \qvec{U_1\, U_2\, U_3\, U_4}$. This state can be written as
\bel{\label{eq split}
  \qvec{\square_p} = \qvec{\underline{\phantom{n}}_{\,\alpha\beta}} \qvec{\sqcap_{\beta\alpha}},
}
with
\bel{
  \qvec{\underline{\phantom{n}}_{\,\alpha\beta}} \equiv \int_G (U_1)_{\alpha\beta} \qvec{U_1}_{\ell_1}, \quad \qvec{\sqcap_{\beta\alpha}} \equiv \int_G (U_2)_{\beta\gamma} (U^{-1}_3)_{\gamma\delta} (U^{-1}_4)_{\delta\alpha} \qvec{U_2}_{\ell_2} \qvec{U_3}_{\ell_3} \qvec{U_4}_{\ell_4}
}
The key point here is that the gauge-invariant state $\qvec{\square_p}$ is written as a gauge-invariant entangled combination of gauge-variant states $\qvec{\underline{\phantom{n}}_{\,\alpha\beta}}$ and $\qvec{\sqcap_{\alpha\beta}}$. Instead of using the single basis element $\qvec{\square_p}\in \H$, we may instead use a larger basis consisting of the $2N^2$ elements $\qvec{\underline{\phantom{n}}_{\,\alpha\beta}}$, $\qvec{\sqcap_{\alpha\beta}} \in \H_0$; as long as we act on it only with gauge-invariant operators and density matrices built out of gauge-invariant states like $\qvec{\square_p}$, we will never ruin gauge invariance, and as an upshot we will be able to cleanly split this enlarged basis into elements in $V$ and elements in $\bar V$. Tracing over the gauge-variant basis elements in $\bar V$ gives a reduced density matrix that maximally mixes the gauge-variant basis elements in $V$. The entropy coming from this density matrix is $2\log N$.

In general, the effect of splitting a flux line of representation $r$ will be to increase the entanglement entropy of the appropriate sector by $\log d_r$. For each piercing of the entanglement edge by a loop with representation $r$, another factor of $\log d_r$ should be added. Keeping this in mind we may now forget all about using gauge-variant basis elements, include this extra entropy $\sum\_{piercings} \log d_r$ by fiat, and pretend that $\H^{(\b k)}$ factorizes into a direct product $\H^{(\b k)}_V \otimes \H^{(\b k)}_{\bar V}$, where $\H_V^{(\b k)}$ contains all gauge invariant states defined purely on $V$ \emph{and} all gauge-variant states represented by open flux lines that end at $\del V$ and that have outgoing flux given by $\b k$. This contribution to the entanglement entropy was identified in \cite{Donnelly:2011hn}, and further discussions of working with gauge-invariant operators on gauge-variant basis states may be found in \cite{Radicevic:2014kqa, Ghosh:2015iwa}.

We now return to constructing the desired reduced density operator. The density operator $\rho$ of a general state may have elements that mix two superselection sectors, but since no gauge-invariant operators in $\A_V$ can change the superselection sector of a given state, these elements of $\rho$ can be set to zero. This way one obtains a block diagonal matrix $\~\rho = \bigoplus_{\b k} p_{\b k}\, \rho^{(\b k)}$ that is sufficient to describe all expectation values of operators in $\A_V$ \cite{Casini:2013rba, Radicevic:2014kqa}. Here $\rho^{(\b k)}$ is a density matrix for states in the $\H^{(\b k)}$ sector and $p_{\b k} \equiv \Tr_{\H^{(\b k)}} \rho$ are c-numbers chosen so that $\rho^{(\b k)}$ has unit trace. By tracing out the $\bar V$ states --- which is possible because we can effectively express $\H^{(\b k)}$ as a direct product --- one obtains the reduced matrix $\rho_V = \bigoplus_{\b k} p_{\b k}\,\rho_V^{(\b k)}$. This is the operator we were after. Its von Neumann entropy is the entanglement entropy we wish to compute. It takes the simple form
\bel{\label{eq S}
  S_V = -\sum_{\b k} p_{\b k} \log p_{\b k} + \sum_{\b k} p_{\b k} S_V^{(\b k)},\quad S_V^{(\b k)} \equiv - \Tr_{\H^{(\b k)}_V}\left(\rho_V^{(\b k)} \log \rho_V^{(\b k)} \right) + \sum\_{piercings}\log d_r.
}
It may be instructive to view this expression as a sum of two types of entropies, a Shannon (or ``classical'') entropy $-\sum_{\b k} p_{\b k}\log p_{\b k}$ that comes from boundary conditions/edge modes, and the average von Neumann entropy $\sum_{\b k} p_{\b k} S^{(\b k)}$ that comes from the entanglement of the interior modes with the exterior.

For states that contain only a few electric flux lines, $\rho^{(\b k)}_V$ will describe pure states and the entanglement entropy will arise solely from the facts that gauge-invariant operators are cut by $\del V$ and that the original state is a superposition of basis elements belonging to different sectors. States of this form are the strong-coupling ground state and its lowest excitations, and we will see that, for them, the Shannon entropy contains the most relevant information about entanglement.

Ground states at weak coupling will feature a superposition of all superselection sectors. If the gauge group is continuous, the superposition will be Gaussian; if it is discrete, the superposition will be uniform. (We will review this in the Appendix.) In the latter case the dominant contribution to the universal parts of the entanglement entropy will come from the $S^{(\b k)}_V$ entropies, while in the former case the Shannon entropy provides corrections comparable to the von Neumann entropy, as has been explicitly demonstrated in \cite{Donnelly:2014gva}. The weak-coupling logarithmic term that we will focus on in Section \ref{sec weak} is unaffected by the presence of the edge modes.

\subsection{Comments}

The procedure described above certainly defines a gauge-invariant quantity that measures entanglement, but a few words are needed to justify calling it \emph{the} entanglement entropy. Implicit choices were made at two junctures in the above discussion:
\begin{description}
  \item[Maximal algebra of observables:] The algebraic approach \cite{Casini:2014aia, Casini:2013rba, Ghosh:2015iwa, Radicevic:2014kqa, Aoki:2015bsa} shows in a particularly clear way that there exist alternative choices for the definition of entanglement entropy. Given a set of links $V$, we may start with any algebra of gauge-invariant observables that are defined only using operators $U_\ell$ and $J_\ell$ with $\ell \in V$. In this paper, we have chosen to work with the maximal algebra $\A_V$ that contains \emph{all} possible operators supported on $V$. Other algebras, such as those with a trivial center (and hence without superselection sectors), will lead to different reduced density operators and different entropies. The maximal algebra is often called the ``electric center'' choice \cite{Casini:2013rba}. \\
      \hspace*{1em}The existence of these choices is by no means unique to gauge theories. A real scalar on a lattice has two operators at each site, $\phi_i$ and its conjugate momentum $\pi_i$. After choosing a set of sites $V$, we still have the freedom to choose an algebra generated by, say, all the $\phi_i$'s  but only some $\pi_i$'s in $V$. Given a state of the scalar field on the entire lattice, the reduced density operator that belongs to this algebra will not be the same as the reduced density operator constructed out of \emph{all} possible $\phi_i$'s and $\pi_i$'s that lie on sites in $V$. The von Neumann entropies of these operators will be different, as well.\\
      \hspace*{1em} The point of view of this paper is that the natural object to study is the maximal algebra of observables in a given region $V$. One reason for this is that the entropy associated to this algebra has a nice interpretation in terms of a replica trick path integral \cite{Ghosh:2015iwa}. Another reason is that this seems to be the object that has been implicitly studied by most other approaches to gauge theory entanglement, and it is for this choice that we recover the familiar notion of topological entanglement entropy. A final reason is that working with the maximal algebra seems to be the approach that was already adopted for theories with matter and no gauge fields. Studying the entropies of non-maximal algebras remains a worthwhile task, and in particular it is of interest to know whether there are measures of entanglement that do not depend on the particular choice of algebra, as long as it is supported on the links in $V$ and not in any subset of $V$ \cite{Casini:2014aia}.
  \item[Gauge theory as a projection:] Another choice we made was viewing the gauge theory as a projection to a $G$-invariant sector of a bigger theory with symmetry group $G$. For instance, the Hilbert space $\H$ of a gauge theory could have been obtained from $\H_0$ as in Kitaev's toric code, by simply positing that the mass of charged states is much greater even than the energy scale set by the lattice spacing \cite{Kitaev:1997wr}. This approach has been contrasted to working with a ``true'' gauge theory, in which there is no physical interpretation of the enlarged Hilbert space $\H_0$ and the factors of $\log d_r$ in \eqref{eq S} are absent \cite{Hung:2015fla}. \\
      \hspace*{1em} Whether one of the above options is more fundamental than the other is a deep question that we will not tackle in depth here; for our purposes this choice is a matter of taste. The present paper takes the former (``projection'') approach because it appears more natural if we are interested in placing the gauge theory on manifolds with nontrivial topology. In the latter approach, Wilson loops along noncontractible cycles would need to be added manually to the set of usual plaquette excitations on flat space, whereas in the projection language the holonomies are automatically included from the start. The dichotomy between the two views of gauge theory seems intimately related to the old question of compact versus noncompact gauge groups: choosing a noncompact gauge group matches the Gaussian fluctuations in the compact theory but misses the topological/nonperturbative effects such as vortices and monopoles, and as a result it typically yields a nonunitary quantum theory unless these defects are added to the theory by fiat.
\end{description}

Even after the above choices are made, we are still left with a variety of methods to calculate the entanglement entropy, as enumerated in Subsection \ref{subsec overview}. We will now comment on the equivalence of these approaches. The discussion of gauge-invariant entanglement of gauge-variant degrees of freedom around eq.~\eqref{eq split} serves to justify the ``embedding approach'' to defining a reduced density operator \cite{Levin:2006zz, Buividovich:2008yv, Donnelly:2011hn, Donnelly:2012st, Donnelly:2014gva, Buividovich:2008gq}. If we embed  the physical Hilbert space into a space where matter degrees of freedom can live on the entanglement edge --- effectively splitting the links into two, as done in \cite{Levin:2006zz} --- we can directly construct the appropriate density operator $\rho_V$ by the usual tracing out procedure. As long as the initial state is gauge-invariant, $\rho_V$ will entangle these edge degrees of freedom so that it computes the correct expectations of gauge-invariant operators in $\A_V$ while giving zero for gauge-variant operators. (In fact, the spaces $\H^{(\b k)}_V$ can be viewed as subsets of such a larger Hilbert space.) Extending this thought, instead of ever working with the physical space $\H$, we can just work with $\H_0$ from the outset and ask for the entropy of the algebra of gauge-invariant operators $\A_V$ acting on $\H_0$; the answer will be the same as for $\H$, and it is given by tracing out the elements of $\H_0$ defined on $\bar V$ \cite{Ghosh:2015iwa}. This tracing out can be expressed by a lattice path integral over all the link configurations, and this connects the calculations in the previous subsection to the ones done by usual replica trick methods. This demonstrates the equivalence of all the methods that have been proposed for calculating entanglement entropy in gauge theories.

We can also connect these lattice computations to the ones done in the continuum limit. Consider a set of links $V$ and focus on a particular site $i \in \del V$. To simplify, pick $i$ so that there is only one link emanating from it that is in $\bar V$. Now consider adding this one link to $V$, thereby creating a new set $V + \delta V$. As long as adding this link has not introduced an entire new plaquette into $V$, the Gauss law guarantees the equality of algebras, $\A_V = \A_{V + \delta V}$, and of the associated entropies $S_V = S_{V + \delta V}$. Thus, as the continuum limit is taken, the boundary of the entangling region is realized as a ``belt'' or ``buffer zone'' of thickness equal to the lattice spacing; the entanglement entropy is not sensitive to whether the set of links $V$ is chosen to contain links in this belt or not. (For example, the link containing the $\b E_3^2$ operator on Fig.~\ref{fig def} can be removed and the link containing the $\bar{\b E}^2_2$ operator can be added to $V$ without affecting the entropy.) Thus, there are equivalence classes of algebras that all have the same entropy, and this explains why it is sensible to draw the entangling edge as a line cutting through links --- if removing a link keeps you in the same equivalence class, then it is meaningless to ask if that link is in $V$ or not, and the entanglement edge might as well be drawn as cutting the link for illustrative purposes. A reasonable conjecture, supported by the results of \cite{Ohmori:2014eia}, is that each equivalence class corresponds to a different entangling edge in the continuum path integral, with classes differing by a single generator corresponding to continuum entanglement edges differing by a single operator insertion along the edge.

We close this section by recalling that there are other choices for boundary conditions that label superselection sectors. (The detailed construction is given in \cite{Radicevic:2014kqa}, and here we just mention the basics.) Each choice corresponds to a commuting set of operators in the ``buffer zone'' of plaquettes around the entangling region, and the entanglement entropy does not depend on this choice. For instance, in $d = 2$, instead of specifying all the electric Casimirs in Fig.~\ref{fig def}, we can specify the values of Wilson loops around the green plaquette, of the Casimirs of the total electric field through the green plaquette, and of electric Casimirs at the remaining boundary sites. A choice that will be particularly useful for studying the weakly coupled regime are the magnetic boundary conditions, where one exclusively works with magnetic operators in the ``buffer zone.'' The superselection sectors are labeled by independent values of $W_p^r$ for all plaquettes $p$ in the buffer zone. For $U(1)$ theory, in the magnetic basis where Wilson loops are diagonalized, the sectors are labeled by a number $w_p = W_p$ for each boundary plaquette. Denoting by $\b w$ the set of all these labels, the entanglement entropy can be expressed analogously to \eqref{eq S} as
\bel{\label{eq S magn}
  S_V = -\sum_{\b w} p_{\b w} \log p_{\b w} + \sum_{\b w} p_{\b w} S_V^{(\b w)}.
}
Below, we will give an explicit example of a calculation using these boundary conditions.

\section{Example: Entanglement at strong coupling}\label{sec strong}

The preceding definitions are most clearly illustrated by studying the strong coupling regime on the lattice. The strong-coupling physics is dominated by the electric term in the Hamiltonian \eqref{H}. The $g = \infty$ ground state, $\qvec{\Omega_{g = \infty}}$, is the state $\qvec \Omega$ introduced in \eqref{def Omega}. Given any region $V$, this state lies in the $\b k = (0, \ldots, 0)$ sector, and its entanglement entropy is
\bel{
  S_V[\Omega_{g = \infty}] = 0.
}
Lattice gauge theories confine at strong coupling, and if $g$ is sufficiently high we expect the confinement scale to be smaller that the lattice spacing. This intuition agrees with the lack of any entanglement structure at distances that can be probed by $S_V$.

As we move away from infinite coupling, the ground state receives corrections from single-plaquette fundamental and antifundamental excitations $\qvec{\square_p}$ and $\qvec{\bar \square_p}$, cf.~eq.~\eqref{def square}. The corrected ground state is found using ordinary perturbation theory. This was done for $G = SU(2)$ in Ref.~\cite{Donnelly:2011hn}.

The leading term in the entanglement entropy comes from the first-order corrected state
\bel{
  \qvec{\Omega_{g \gg 1}} = \left(1 - \frac{N_P}{\lambda^2}\right) \qvec \Omega + \frac1\lambda \sum_p \big(\qvec{\square_p} + \qvec{\bar\square_p}\big),
}
where $N_P$ is the (finite) number of plaquettes on the lattice, and the effective coupling is $\lambda \equiv s g^4N$, with $s$ being the number of links on a plaquette. The $O(1/\lambda^2)$ term serves to normalize the state. It can be shown that other $1/\lambda^2$ corrections that would come from second-order perturbation theory do not contribute to the entropy at leading order. If $\lambda^2 \sim N_P$, we need to go to higher orders in perturbation theory to obtain properly normalized states; to avoid technical complications we will assume that $\lambda^2 \gg N_P$.

After picking a subset of links $V$, applying eq.~\eqref{eq S} is straightforward. The starting density matrix is $\rho = \qvec{\Omega_{g \gg 1}}\qvecconj{\Omega_{g \gg 1}}$. States $\qvec\Omega$, $\qvec{\square_p}$, and $\qvec{\bar\square_p}$ all belong to the same sector, $\b k = (0, \ldots, 0)$, when the plaquette $p$ has all its links in $V$ or all its links in $\bar V$. Plaquette excitations associated to $p$'s that are orthogonal to $\del V$, i.e.~that have links both in $V$ and $\bar V$, lie in different sectors. Let $\del V_\perp$ be the set of such plaquettes. For each $p \in \del V_\perp$, $\qvec{\square_p}$ and $\qvec{\bar\square_p}$ belong to the sector labeled by a nonzero $\b k$. The block-diagonal matrix $\~\rho$ takes the form
\bel{
  \~\rho = \qvec\Psi \qvecconj\Psi + \frac1\lambda \sum_{p \in \del V_\perp} \big(\qvec{\square_{p}} + \qvec{\bar\square_{p}}\big) \big(\qvecconj{\square_{p}} + \qvecconj{\bar\square_{p}}\big),
}
where we have defined the auxiliary state
\bel{
  \qvec\Psi \equiv \left(1 - \frac{N_P}{\lambda^2}\right) \qvec \Omega + \frac1\lambda \sum\nolimits'_p \big(\qvec{\square_p} + \qvec{\bar\square_p}\big)
}
with $\sum'_p$ denoting the sum over all plaquettes that are not perpendicular to $\del V$. In each sector we have a pure state, so the entanglement entropy \eqref{eq S} only receives contributions from the Shannon entropy of the $p_{\b k}$'s and from the cuts through gauge invariant states (with two $d_r = N$ piercings in each sector where $\b k \neq 0$):
\algnl{\notag
  S_V[\Omega_{g \gg 1}]
  &= -\left(1 - \frac{2|\del V_\perp|}{\lambda^2}\right)\log\left(1 - \frac{2|\del V_\perp|}{\lambda^2}\right) - \frac{2|\del V_\perp|}{\lambda^2} \log\frac2{\lambda^2} + \frac{2|\del V_\perp|}{\lambda^2} \log N^2 + O\left(\frac1{\lambda^2}\right)\\ \label{S gg 1}
  &= \frac{2|\del V_\perp|}{\lambda^2} \left(1 + \log \frac{\lambda^2N^2 }2\right) + O\left(\frac1{\lambda^2}\right).
}
This is the large-$N$ generalization of the entropy obtained in \cite{Donnelly:2011hn}. At strong coupling the entropy thus vanishes as $S_V \sim |\del V_\perp|\dfrac{\log g^2 N}{g^8 N^2}$, and the planar limit only accelerates the vanishing.

\section{Entanglement at weak coupling}\label{sec weak}

\subsection{The weak coupling limit}

We now focus on the small-$g$ limit of the Hamiltonian \eqref{H}. The conceptually simplest way to proceed is to fix the axial gauge. It may be useful to spell out this procedure in the context of our work. In the Hamiltonian formalism, going to axial gauge amounts to replacing the physical Hilbert space $\H$ with an isomorphic space $\H^\star \subset \H_0$ that is spanned by states $\qvec U_\ell^\star$ located on a particular subset of all links on the lattice, the ``living'' links. In this nomenclature, ``dead'' links are the ones on which we can use a gauge transformation to set $U_\ell = \1$, and all remaining links are ``living.'' On a hypercubic lattice one typically chooses all links in one direction (say, along the $x$-axis) to be dead, then one picks a fixed-$x$ slice of the lattice and kills off all $y$-directed links on this slice, and so on through all the directions of the lattice. Once gauge freedom is completely exhausted, we are left with $N_P$ living links (one per plaquette) that directly correspond to states that span the physical Hilbert space. The states on living links are now allowed to take \emph{any} value without restrictions.

It is important to stress that all gauge-invariant operators that act on $\H$ can still be defined on $\H^\star$. The Gauss law allows us to express electric operators on dead links as functions of electric operators on living links, and the gauge-fixing condition instructs us to just set $U_\ell = \1$ on all dead links $\ell$ that appear in magnetic operators. Thus no observables are dropped by the gauge-fixing.

At weak coupling, Yang-Mills theory is either in the Coulomb phase, or it confines at a length scale that grows exponentially with $1/g^2$. Thus, at sufficiently small distances and couplings the theory will always appear to be in the Coulomb phase. (See Appendix \ref{app weak} for a more precise justification of this statement.) The ground state $\qvec{\trm{Coulomb}}$ is a direct product of $\trm{dim}(G)$ ground states of identical, free, noncompact photons. The total entanglement entropy in this state is
\bel{
  S_V[\trm{Coulomb}] = \dim(G)\ S_V\^{(photon)}.
}
This entanglement entropy scales as $S_V \sim N^2$, in sharp contrast with the swift $\sim (\log N)/N^2$ vanishing of $S_V$ at strong coupling and large $N$, as found in eq.~\eqref{S gg 1}.

The entanglement entropy of a single noncompact photon, $S_V\^{(photon)}$, can be computed using methods developed in Section \ref{sec prelim}.  In $d = 2$, this quantity has been investigated numerically in the gauge theory \cite{Casini:2014aia}, and analytically in the dual scalar theory \cite{Agon:2013iva}. We now present a direct, analytic gauge theory calculation that shows that the entanglement entropy has the form
\bel{\label{ad S}
  S_V\^{(photon)} = (d - 1) \left(S_V\^{(scalar)} + \Delta S(g)\right) + \trm{corrections},
}
where $S_V\^{(scalar)}$ is the entanglement entropy of a massless scalar, and
\bel{\label{ad Delta S}
  \Delta S(g) = \frac12 \log\left(g^2 |\del V|\right)
}
is a ubiquitous coupling-dependent term that arises because noncompact photons, just like Nambu-Goldstone bosons, lack a zero mode; in other words, this term comes about because weakly coupled gauge fields (in the right gauge) can be realized as $d - 1$ scalar fields with the identification $\phi(x) \equiv \phi(x) + \eps$. The ``corrections'' above are primarily terms that arise due to the presence of the edge modes; they will not be our concern in this paper, but we will outline how they are computed. The $\Delta S$ terms exist in all dimensions but the $d = 2$ ones have a particular significance, as we will discuss below.

\subsection{Warm-up: The $O(M)$ model}

As mentioned above, $\Delta S(g)$ is a term that can be found in systems with Nambu-Goldstone modes. Before calculating in the gauge theory, we now review how such ubiquitous logarithmic terms arise in symmetry-breaking ground states of nonlinear $\sigma$-models \cite{Metlitski:2011}. Consider a continuum $O(M)$ model of radius $\sigma$ in $d$-dimensional flat space. The action is
\bel{
  S = \sigma^2 \int \d t\, \d^d \b x \left(\frac12 (\del_t \vec n)^2 - \frac12 (\bfnabla \vec n)^2 \right),\quad \vec n^2 = 1.
}
We wish to study the physics of $\vec n$ configurations that are all very close to a particular direction $\vec n_0$. This is a free theory, so given a region $V$, the ground state reduced density matrix for these fluctuations can be explicitly calculated to be
\algnl{\notag
  \rho_V[\vec n, \vec n']
  &\propto \exp\left\{ - \frac{\sigma^2}8 \int_V \d^d \b x\, \d^d \b y \left(\vec n(\b x) - \vec n'(\b x) \right) Q(\b x, \b y) \left(\vec n(\b y) - \vec n'(\b y) \right) \right\} \times \\ \label{rho n nprim}
  &\qquad \times \exp\left\{ - \frac{\sigma^2}8 \int_V \d^d \b x\, \d^d \b y \left(\vec n(\b x) + \vec n'(\b x) \right) \Delta(\b x, \b y) \left(\vec n(\b y) + \vec n'(\b y) \right) \right\},
}
where
\bel{\label{def Q}
  Q(\b x, \b y) \equiv \frac2{V\_{system}} \sum_{\b k} |\b k| e^{i\b k (\b x - \b y)}
}
specifies the Hamiltonian $\frac12 \int \d^d \b x\, \d^d \b y\, \delta \vec n(\b x) Q(\b x, \b y) \delta \vec n(\b y) = \sum_{\b k} |\b k| \delta \vec n_{\b k}\+ \delta \vec n_{\b k}$ of the fluctuations $\delta \vec n = \vec n - \vec n_0$, and $\Delta(\b x, \b y)$ is a complicated kernel that arises after integrating out the modes outside $V$ and whose form we will not need. Throughout this derivation, it is assumed that fluctuations are small; the self-consistency of this assumption must be checked by computing the size of fluctuations \`a la Coleman-Weinberg. (For instance, from the Mermin-Wagner theorem we know that in $d = 1$ there will be no symmetry-breaking phase with small fluctuations.)

We now decompose $\vec n$ into a soft (``zero'') mode $\vec n_z$ and fluctuations $\chi_a$, $a = 1, \ldots, M - 1$, defined through\footnote{The presence of $\sigma$ in this expansion is optional and just makes the $\chi$'s be dimensionful scalars that take values much smaller than $\sigma$.}
\bel{
  \vec n \equiv \vec n_z \sqrt{1 - \frac{\chi_a \chi_a}{\sigma^2}} + \frac{\vec e_a \chi_a}\sigma,\quad \vec e_a \cdot \vec n_z = 0,\quad \vec e_a \cdot \vec e_b = \delta_{ab},\quad \int_V \d^d \b x\, \d^d \b y\ Q(\b x, \b y) \chi_a(\b y) = 0.
}
The last condition justifies the name ``soft'' or ``zero'' mode for $\vec n_z$, as it makes sure that $\vec n_z$ contains all information about modes whose wavelengths are greater than the size of $V$. In these new variables, the reduced density matrix takes the form
\bel{\label{rho V}
  \rho_V[\vec n, \vec n'] \propto \exp\left\{-\frac I2 \left(\vec n_z - \vec n_z'\right)^2\right\}\ \~\rho_V[\chi_a, \chi'_a],\quad I \equiv \frac{\sigma^2}4 \int_V \d^d \b x\, \d^d \b y\ Q(\b x, \b y),
}
where $\~\rho_V$ looks just like $\rho_V$ in eq.~\eqref{rho n nprim}, but with compact fields $\vec n$ replaced by noncompact fields $\chi_a$. The dimensionless parameter $I$ can be calculated straight from eq.~\eqref{def Q} and equals
\bel{\label{eq I}
  I = \frac{\sigma^2}{2\pi} |\del V| \left(\log \frac ra + \ldots\right),
}
where $a$ is the lattice spacing/UV cutoff, and $r$ is the IR cutoff of the integral over $V$ (i.e.~$r$ is the linear size of $V$). The dots represent corrections to the logarithm that are determined from the exact shape of $\del V$. For any macroscopic region with $r \gg a$ we have $I \gg 1$, meaning that the density matrix for $\vec n_z$ is close to the identity.

The reduced density operator $\rho_V$ has a very particular form. As an operator on the subspace associated to the soft mode $\vec n_z$, $\rho_V$ has matrix elements of the form $e^{-I (\vec n_z - \vec n_z')^2}$ and so, in operator form at $I \gg 1$, $\rho_V \sim e^{- \vec J^2/2I}\~\rho_V$, where $\vec J$ is the spin operator conjugate to $\vec n$. The exponent has the characteristic form of the ``Anderson tower of states'' Hamiltonian that describes the small gaps between vacua of a system with continuous symmetry breaking. This is a crucial observation: the symmetry-breaking ground state of the $O(M)$ model is invariant under global rotations and does not depend on the zero mode of the system, and this lack of a zero mode forces the softest mode of the subsystem to have a very specific, $\sigma$-dependent entanglement spectrum with the environment.

The eigenvalues of the ``tower of states'' modular Hamiltonian scale as $1/I \sim \left( \sigma^2 r^{d - 1} \log(r/a)\right)^{-1}$. Elements of $\~\rho_V$ have the form $e^{- Q (\chi - \chi')^2 - \Delta (\chi + \chi')^2}$. The presence of the $(\chi + \chi')^2$ term in the exponent makes this matrix qualitatively different from the soft mode density matrix, and indeed, in operator form $\~\rho_V$ is an exponential of an SHO Hamiltonian whose energy levels scale as $1/\log(r/a)$ and are independent of $\sigma$ \cite{Metlitski:2011}. The modular Hamiltonian thus splits into two parts --- a quantum rotor that describes the soft mode states and an SHO that describes the other modes --- and only the first part still carries $\sigma$-dependence. It is through this part of the density matrix that one recovers the coupling-dependent universal term announced in eq.~\eqref{ad Delta S},
\bel{\label{eq Delta S rho}
  \Delta S(\sigma) = \frac{M - 1}2 \log(\sigma^2 |\del V|),
}
easily observed in the von Neumann entropy of the operator $e^{-\vec J^2/2I}$. The remaining terms from the soft mode entanglement entropy, such as various constants and a $\log\log(r/a)$ term, are also found in the entanglement entropy of the $\chi^a$ modes, and are not as ubiquitous as $\Delta S(g)$. The resummation of these terms is beyond the scope of this work, but \cite{Metlitski:2011} have argued that this resummation gives the usual area law term and the accompanying subleading corrections.

In $d = 2$, the renormalized entanglement entropy for a circle is $F(r) = r S'(r) - S(r)$, and therefore the ubiquitous coupling-dependent term in two spatial dimensions is
\bel{
  \Delta F(\sigma) = - \frac{M - 1}2 \log(\sigma^2 r).
}
The remaining contribution to $F(r)$ does not depend on $\sigma$. As we flow to the IR and $\sigma$ increases, $\Delta F$ (and hence the entire $F$ quantity) decreases, as per the $F$-theorem.

\subsection{Gauge theory}

We now return to the gauge theory case and repeat the same analysis. The first order of business is to find the ground state wavefunction and then reduce the density matrix in order to get the analog of eq.~\eqref{rho n nprim}. The gauge-fixed Hamiltonian is
\bel{
  H =  \frac{g^2}2 \sum_{\ell} \b J^2_\ell + \frac1{2g^2} \sum_p \left(2N - \Tr W_p - \Tr \bar W_p\right),
}
where, on dead links, we set $A_\ell = 0$ and express $\b J^2_\ell$ in terms of living links using the Gauss law. The ground-state wavefunction can be found by expanding in small fluctuations around $U = \1$, diagonalizing the Hamiltonian, and determining the usual SHO ground state for each of the eigenmodes.

After using the Gauss law and expanding in fluctuations $A_\ell^a$, the Hamiltonian takes the form
\bel{
  H = -\frac12 g^2 {\sum_{\ell,\,\ell'}}^\star K_{\ell\ell'} \fder{}{A^a_\ell} \fder{}{A^a_{\ell'}} + \frac1{2g^2} {\sum_{\ell,\, \ell'}}^\star M_{\ell\ell'} A^a_\ell A^a_{\ell'}.
}
The ground state wavefunction is
\bel{\label{eq psi A}
  \psi[A] = \left(\frac{\det Q}{(\pi g^2)^{N_P}}\right)^{\dim(G)/4} e^{- \frac1{2g^2} \sum_{\ell,\,\ell'}^\star Q_{\ell\ell'} A_\ell^a A_{\ell'}^a},
}
where $\avg{A^a_\ell A^b_{\ell'}} \propto g^2 \delta^{ab} \left(Q^{-1}\right)_{\ell\ell'}$ is the gluon propagator in this gauge, and
\bel{
  Q \equiv K^{-1/2} \left(K^{1/2}MK^{1/2}\right)^{1/2} K^{-1/2}.
}
This expression for the propagator and for the ground state is correct as long as there are no nonperturbative effects that cause the propagator to change at large distances. Such effects can indeed exist, e.g.~they are present in theories with confinement via monopoles, but as long as we work with small enough coupling and at small enough distances, the above ansatz for the ground state will be correct. We will comment more on this issue below.

As an example, in $d = 2$ and at large distances compared to the lattice spacing, the matrix $Q$ is given by
\bel{\label{def Qxy}
  Q_{\ell \ell'} \propto \sum_{\b k} \frac{k_1^2}{\sqrt{k^2_1 + k_2^2}} e^{i \b k \cdot (\b x - \b x')},
}
where $\b x$ and $\b x'$ are coordinates associated to links $\ell$ and $\ell'$. (At smaller distances one has to replace $k_i$ by $2\sin(k_i/2)$ to get the correct lattice propagator.) As in the $O(M)$ model, the wavefunction in terms of the original variables is invariant under global shifts in $A^a_\ell$, which is seen from the fact that $\sum_{\ell, \ell'}^\star Q_{\ell\ell'} = 0$.

The density matrix corresponding to the state \eqref{eq psi A},
\bel{
  \rho[A, A'] = \psi^*[A] \psi[A'],
}
must be reduced following the prescription appropriate to the gauge theory. This is where one uses magnetic boundary conditions and eq.~\eqref{eq S magn}. If one were to ignore the presence of edge modes and the need to decompose the density matrix into subsectors, one would get a gauge-dependent result: states with a flux loop intersecting the edge $\del V$ could be in principle represented as excitations in $V$ or outside $V$, depending on the gauge choice, and the edge modes are the gauge-invariant way to keep track of flux loops that enter and exit the entangling region.

Consider the case of one photon. The superselection sectors are labeled by the values of Wilson loops in the ``buffer zone'' around the region $V$; these live on plaquettes that have links both in $V$ and in $\bar V$, such as the green plaquette in Fig.~\ref{fig def}. In axial gauge, a  Wilson loop on a plaquette is equal to the difference of $A_\ell$'s on the two living links belonging to the plaquette. One superselection sector thus consists of all field configurations $A_\ell$ that satisfy $A_\ell - A_{\ell'} = w_p$ for two living links $\ell$, $\ell'$ in the edge plaquette $p$. This sector is labeled by the collection of all the $w_p$'s, denoted $\b w$, and we will call the sector $\H^{(\b w)}$. We can now work sector by sector, defining the block density matrix in a given sector as $p_{\b w} \rho^{(\b w)}[A, A'] = \rho[A, A']$, where $A$ and $A'$ both belong to the sector labeled by $\b w$. The normalization constants are defined as $p_{\b w} = \Tr_{\H^{(\b w)}} \rho$, so the operators $\rho^{(\b w)}$ have unit trace. Now we can trace out the degrees of freedom on living links in $\bar V$, getting the reduced operator $\rho_V^{(\b w)}$ whose von Neumann entropy figures in eq.~\eqref{eq S magn}.

For a general gauge group, the superselection sectors are labeled by a set of numbers $w_p^a$ for each gluon. Given a sector $\b w$, we may now repeat the steps of the previous section and extract the density matrix for the softest mode in $V$. This density matrix will once again be the exponential of a ``tower of states'' Hamiltonian with prefactor proportional to $I = \frac1{g^2}\int_V \d^d \b x\, \d^d\b y\,   Q_{\b x \b y}$, just like in eq.~\eqref{rho V}. The relevant logarithmic term in the von Neumann entropy of the soft mode in this sector will be $\Delta S_{\b w} = \frac{\dim(G)}2 \log I$. This term will not depend on $\b w$ because the soft mode coupling only depends on $Q_{\ell \ell'}$ and $g^2$, so the logarithmic terms in the full entropy add up to $\Delta S(g) = \sum_{\b w} p_{\b w} \Delta S_{\b w}(g) = \frac{\dim(G)}2 \log I$ due to the normalization condition $\sum_{\b w} p_{\b w} = 1$.

We will prove that $g^2 I = |\del V|\, f\left(|\del V|/a^{d-1}\right)$, where $f(x)$ is a slowly varying function, e.g.~a log. For notational simplicity, let us focus on $d = 2$ with $V$ being a region of linear size $r$. We now wish to show that $g^2 I \propto r f(r/a)$. The most direct argument goes as follows. If we let $Q_{\b x \b y} = Q(\b x - \b y)$, the function $Q(\b x)$ will behave as $Q(\b x) \sim x_1^2 / |\b x|^5$ at $|\b x| \gg a$, assuming that the full system size is much larger than any other scale. Any divergences that appear in $I$ will come from the region when $\b x$ and $\b y$ are close to each other. Up to constant prefactors, the divergent pieces will be the same as in $\int_V \d^2 \b x\, \d^2 \b y\, |\b x - \b y|^{-3}$. After integrating out $\b y$ from this integral, we will be left with
\bel{
  I = \frac1{g^2} \int_V \d^2\b x \left(\frac ca + f_0(\b x) + a f_1(\b x) + \ldots \right).
}
By dimensional analysis and symmetries, $c$ must not depend on $\b x$ and $f_0(\b x)$ cannot have a divergence worse than $1/|\b x - \b x_0|$ for some set of points $\b x_0 \in V$. Integrating over $\b x$ will give
\bel{
  I = \frac1{g^2}\left(c\, \frac{|V|}{a} + r f(r/a) + O(a)\right),
}
where $f$ contains nothing worse than a logarithmic divergence. Now, if $V$ had been the entire system and if we had used the exact lattice propagator, $I$ would have been zero by the definition of $Q_{\b x \b y}$ in \eqref{def Qxy}. This is consistent with the volume term $c |V|/a$ found above only if $c$ renormalizes to zero when using the exact propagator. This proves that $g^2I = r f(r/a)$ in a continuum description.

We have checked this numerically for fixed superselection sectors on small (up to $100\times 100$), $d = 2$, square lattices. A detailed numerical analysis analogous to \cite{Luitz:2015, Kulchytskyy:2015} could be used to determine the exact form of $f(r/a)$ and therefore the remaining terms in the entanglement entropy.

An analytic argument in favor of the $I \sim |\del V|$ scaling in any $d$ goes as follows. Instead of doing the microscopic calculation outlined above, let us construct a toy model. We wish to study slow fluctuations of the gauge field in the region $V$ and to replace the entire field $U_{\ell}$, $\ell \in V$, with a single effective degree of freedom $U$. The effective Lagrangian for $U$ and its environment $W$ is
\bel{\label{def L toy}
  L = \frac12 c_U \Tr\left(\del U^{-1} \del U\right) + \frac12 c_W \Tr\left(\del W^{-1} \del W\right) - \frac J2 \left[2N - \Tr\left(U^{-1} W\right) - \Tr\left(U W^{-1}\right)\right],
}
The average coordinate $U$ in this region has coupling $c_U \sim |V| \sim r^d$, where $r$ is the characteristic size of $V$. If $V$ is much smaller than the entire system, the relative weight $c_r = c_U c_W/(c_W + c_U)$ will also scale like $r^d$. On the other hand, the coupling $J$ in \eqref{def L toy} is the characteristic spin-wave coupling of a system to its boundary conditions that scales as $J \sim r^{d - 2}$ for $d \geq 2$ \cite{Metlitski:2011}. Thus, the coupling of the soft (and only) mode in the effective, toy-model description is found to scale as $\xi^2 \sim 1/\sqrt{c_r J} \sim 1/r^{d - 1}$. The ground state wavefunction of the above Lagrangian is
\bel{
  \psi(U, W) = \frac1{\sqrt{\trm{Vol}(G)}} \frac1{(\pi \xi^2)^{\dim(G)/4}}  e^{-\frac1{2\xi^2}\left(2N  - \Tr(U^{-1} W) - \Tr(U W^{-1}) \right)},
}
or
\bel{
  \psi(A) = \frac1{(\pi \xi^2)^{\dim(G)/4}} e^{- (A^a)^2/2\xi^2},
}
in terms of the small fluctuations defined as $U = W e^{i A^a T^a}$. Thus, the soft mode coupling should be $I = 1/\xi^2 \sim |\del V|$. This argument does not prevent possible multiplicative $\log r$ terms in $I$ since the scaling behavior is just captured by the leading powers of $r$, but this is good enough for our purposes.

Putting everything together, we can conclude that the entanglement entropy has a coupling-dependent term of the form
\bel{\label{eq Delta S g}
  \Delta S(g) = \frac{\dim(G)}2 \log\frac{|\del V|}{g^2}.
}
In the continuum limit in $d = 2$, however, only the combination $g^2 |\del V|$ remains finite, so by writing $\log (|\del V|/g^2) = \log(g^2 |\del V|) - 2 \log g^2$ we can extract the term in the coupling-dependent piece with a regular continuum limit,
\bel{\label{eq Delta S e}
  \Delta S(e) = \frac{\dim(G)}2 \log(e^2 r),
}
where $e$ is the continuum coupling and $r$ is the linear size of the region $V$. This is the advertised result \eqref{eq Delta S ad}. Comparing it to other terms in the universal part of the $d = 2$ entanglement entropy, we see that $\Delta S(e)$ at small $e^2 r$ is parametrically larger than the leading universal term (the $F$ term) which takes on $e$-independent values of order unity.

What happens for $d \geq 3$? The coupling-dependent piece \eqref{eq Delta S g} still has the same form. The continuum coupling is $e^2 = g^2 a^{d - 3}$,
meaning that the finite coupling-dependent piece of the entropy is
\bel{
  \Delta S (e) = \left\{
                          \begin{array}{ll}
                            \dim(G) \log e^2, & \hbox{$d = 3$;} \\
                            \frac12 (d - 1)\, \dim(G) \, \log \left(e^{\frac2{3 - d}} r\right), & \hbox{$d > 3$.}
                          \end{array}
                        \right.
}
Thus, in $d = 3$ the log term from the soft modes gets ``contaminated'' with universal constant terms from the other modes. Similar contamination of $\Delta S$ with $\log(r/a)$ terms will happen for other odd $d$. For even values of $d$ a ubiquitous $\Delta S$ term does exist as long as the weakly coupled ground state still displays symmetry breaking, but for $d > 2$ it will not be the dominant universal piece. In all dimensions, we find that $\Delta S$ is proportional to $d - 1$, the number of independent scalar degrees of freedom contained in a gauge field --- this is the origin of the $d- 1$ prefactor in eq.~\eqref{ad S}.

We will not give any specifics on the ``correction'' terms in eq.~\eqref{ad S}, but now we see that they come from the from the $\log g^2$ piece and the subleading behavior of the soft mode coupling $I$, from the von Neumann entropies of all the other modes that do not depend on $g^2$, and from the Shannon entropy of the edge modes which will also depend on $\log g^2$, as per the definition of the $p_{\b w}$'s. The latter was shown to provide a contribution comparable to the one of a single scalar in \cite{Donnelly:2014gva}. None of these entropies can contain the combination $\log\left(e^2 r\right)$, however; the term we have found is indeed ubiquitous in $d = 2$.\footnote{In $d = 3$ this lack of zero mode will shift the coefficient of the $\log\frac ra$ term in the entanglement entropy by the shape-independent constant $\dim(G)$. In this paper we have not discussed the signature of this weak-coupling entanglement in relation to the usual universal $\log\frac ra$ terms in odd $d$, but further discussion on this subject can be found in Ref.~\cite{Pretko:2015zva} that appeared after the first version of this paper.}

Finally, we emphasize that just like in the $O(M)$ model of the previous section, the above calculation on its own gives precious little insight into when the weak coupling, symmetry-breaking description is valid. The result above should thus be interpreted to mean that the Coulomb phase of a sufficiently weakly coupled gauge theory will always have this logarithmic term present, but the presence of the Coulomb phase must be determined by other considerations.\footnote{As we will comment below, nonperturbative effects may wash out this logarithmic term. This is precisely what happens for the $d = 2$ $U(1)$ gauge theory and the $O(2)$ model as one goes deeper into the IR. We make the plausible assumption that for any $d$ and $G$ there exists a coupling regime at which the theory looks deconfined at small enough length scales.}  It remains to be seen whether the presence of $\Delta S$ terms in the entanglement entropy of a gauge theory always implies that the theory is in a Coulomb phase.

\subsection{Related results in $d = 2$ Maxwell theory}

It is instructive to review how other calculations relate to the $d = 2$ result, eq.~\eqref{eq Delta S e}, for gauge group $G = U(1)$. This case has been the subject of much study, both due to the simplicity and richness of the gauge theory (see Appendix \ref{app maxwell}), and because it is possible to dualize the theory to a scalar one (see Appendix \ref{app dual}; invariance of the entanglement entropy under dualities will be discussed in the Conclusion). In particular, instead of the above Hamiltonian analysis, it is possible to carry out an explicit replica-trick calculation in the dual scalar theory, and indeed the entanglement entropy of the compact Maxwell theory on a disk of radius $r$ was found to contain a $\frac12 \log e^2r$ term when $e^2r \ll 1$, with the symmetry-breaking description becoming invalid at $e^2 r \gg 1$ \cite{Agon:2013iva}. (The path integral method has also been used to reproduce the log term in \cite{Metlitski:2011}, although the validity of the calculation at large $e^2 r$ was not the focus of that work.) The disappearance of $\Delta S(e)$ in the IR, when $e^2 r \gg 1$ but still below the confinement scale, is realized as a nonperturbative effect in the path integral language. It is less clear how this effect is realized directly in our Hamiltonian gauge theory analysis; it is related to the fact that the free photon propagator, $Q_{\ell \ell'}$, starts changing at distances much greater than $1/e^2$, invalidating the initial expression for the wavefunction $\psi[A]$ in \eqref{eq psi A}. A detailed understanding of IR effects on the $\Delta S$ term for various $d$ and $G$ remains a topic for future work.

A less direct reproduction of the same result appears in \cite{Klebanov:2011td}, where the renormalized free energy of compact Maxwell theory on a three-sphere of radius $r$ was found to contain a term $-\frac12 \log e^2 r$, and the presence of this term was linked to constant gauge transformations/zero modes on the sphere. While this free energy is very closely related to the renormalized entanglement entropy $F(r)$, these quantities do not match for nonconformal theories. It is interesting that the universal logarithmic term does seem to match; this suggests that similar universal terms might be extractable by performing free energy calculations on spheres.

A numerical lattice calculation for $F(r)$ of a noncompact photon \cite{Casini:2014aia} has yielded the same kind of term, $-\frac12 \log(r \mu)$, for an undetermined, UV-independent $\mu$, by computing the entropy of the ``truncated scalar algebra,'' i.e.~of the operator algebra containing only derivatives of a free scalar field. Working with the truncated algebra is equivalent to working with Nambu-Goldstones/noncompact photons, i.e.~scalars that are simply missing a zero mode. Our analysis, moreover, shows that the constant $\mu$ is determined by the choice of normalization of the gauge field --- while the algebra of observables is invariant under field rescalings (since these are canonical transformations), the entanglement entropy is not \cite{Casini:2014aia}, and hence the operator algebra should be supplemented by a rule to fix this scaling ambiguity. Physically, fixing this scaling amounts to choosing the compact theory whose gauge-fixing or spontaneous symmetry breakdown gives the needed theory of Nambu-Goldstones.

Finally, we point out that the logarithmic term found above is analogous to the topological entanglement entropy found in systems with discrete Abelian gauge groups. Consider the $\Z_2$ gauge theory in $d = 2$. Its entanglement entropy famously contains the universal term $-\log 2$, the topological entanglement entropy \cite{Levin:2006zz, Kitaev:2005dm}. The dual of this gauge theory is the Ising model defined up to a global spin flip. On its own, the Ising model has no topological entanglement entropy; the $-\log 2$ must come from the global $\Z_2$ ambiguity. This ambiguity can be viewed as the omission of the zero mode in the Ising model, in analogy with the $U(1)$ case we studied in detail. We will further comment on this in the Conclusion.

\section{Conclusion}

This paper has presented a computation of entanglement entropy directly in Yang-Mills gauge theory on a lattice. In particular, we have provided a transparent connection between logarithmic terms in the entanglement entropy and the lack of the zero mode in the theory; the arguments in this paper complement the results of papers \cite{Agon:2013iva, Metlitski:2011} from a gauge theory point of view. Our results are particularly significant in $d = 2$, where the logarithmic terms presented here dominate the universal part of entanglement entropy at weak coupling. We now review several points of interest that may warrant further work.

In Section \ref{sec strong} we have shown that at strong lattice coupling the entanglement entropy vanishes as $(\log N)/N^2$ in the planar limit. Conversely, in Section \ref{sec weak} we have shown that at sufficiently weak coupling the entropy scales as $N^2$. The entanglement entropy thus jumps by an infinite amount as the gauge coupling is dialed from strong to weak over a finite interval; this establishes the presence of a lattice phase transition in the planar limit. This result agrees with previous lattice studies \cite{Narayanan:2003fc} and establishes the entanglement entropy as a good order parameter for confinement that can be calculated directly in the gauge theory. The grand prize --- proving the presence of the phase transition at small enough lattice couplings such that the continuum limit is applicable, as done holographically in \cite{Klebanov:2007ws} --- remains beyond our abilities for now.

Reference \cite{Agon:2013iva} has already shed significant light on the origin of logarithmic terms in the $d = 2$ Maxwell theory, but it has nevertheless relied heavily on the Maxwell-scalar duality. How invariant is the entanglement entropy under such dualities? Our calculation has shown that the logarithmic term is present on both sides of the duality. It would be of great interest to understand whether all universal terms are preserved under all dualities of Kramers-Wannier type. Progress on this front is most easily accessible by studying the Ising model and its dual $\Z_2$ gauge theory in $d = 2$. Here it is possible (and easy) to track how the maximal algebra $\A_V$ in a region $V$ of the gauge theory maps across the duality; the result is that $\A_V$ maps to an algebra $\~\A_{\~V}$ on a set of sites $\~V$ on the dual lattice, with $\~\A_{\~V}$ generated by operators $\sigma^x_i$ for $i \in \~V$ and $\sigma^z_i$ for $i \in \trm{Int}\~V = \~V - \del \~V$. In other words, the maximal algebra in the gauge theory does not map to the maximal algebra in the dual theory. (Repeating this analysis in $d = 3$ would show that the maximal algebra, i.e.~the electric center choice, maps to the algebra known as the magnetic center choice in the dual gauge theory.) In $d = 2$, this is reassuring, as the entanglement entropy of the weakly coupled gauge theory could not be equal to the usual entanglement entropy of the strongly coupled Ising model, which lacks the usual area law due to disorder at strong coupling. Fleshing out this duality will be the subject of a future publication.

In a similar vein, at several points we have alluded at an incomplete dictionary between alternative algebra choices and operator insertions in the continuum. Writing this correspondence has been initiated by \cite{Ohmori:2014eia} for the Ising model in $d = 1$. Continuing this program for other theories, both gauge and pure matter, would be a fruitful task.

The analysis in this paper has relied heavily on lattice gauge theory techniques. It would be of interest to develop a continuum approach that takes into account all the subtleties that come with a gauge theory but that does not require working directly with the lattice. Initial steps have been outlined in \cite{Donnelly:2014gva}, and following this program through could lead to a versatile definition of entanglement entropy that can be more readily connected to path integral calculations using the replica trick.

Finally, this paper deals with a rather vast, formal topic: defining entanglement entropy in a theory with nonlocal degrees of freedom. Extending the kind of analysis given in this paper to other theories with gauge constraints, in particular to gravity, would be extremely interesting and is already a subject of investigation \cite{Donnelly:2015hta}.

\section*{Acknowledgments}

I would like to thank Steve Shenker for his continuous support and many insightful ideas and questions. It is also a pleasure to thank Sinya Aoki, Shamik Banerjee, Horacio Casini, William Donnelly, Marina Huerta, Chao-Ming Jian, Edward Mazenc, Kantaro Ohmori, Masahiro Nozaki, Xiao-Liang Qi, Matt Roberts, Lenny Susskind, Yuji Tachikawa, Sandip Trivedi, and members of many audiences that have heard preliminary versions of this work; the questions and comments raised by all these people have helped direct and sharpen the arguments presented here. Finally, I thank Lina Wu for alerting me to a mistake in the original manuscript. The author is supported by a Stanford Graduate Fellowship.

\appendix

\section{Weak coupling details} \label{app weak}

In this section we start from the gauge-fixed Kogut-Susskind Hamiltonian and analyze the weak coupling limit more thoroughly, justifying the statement that the ground state looks like $\dim(G)$ decoupled photons. Let us start by asking what happens at $g = 0$. The na\"ive (and wrong/incomplete) answer is simple: the electric part of the Hamiltonian disappears and the ground state is an eigenstate of the position operators, defined in \eqref{pos mom ops}, such that $W_p = \overline W_p = N$ on each plaquette. Since our lattice is topologically trivial, the only axial gauge configuration that satisfies this has $U = \1$ on all living links. This is the analog of the topological state $\qvec{\trm{topo}}$ found in $\Z_\kappa$ gauge theories, e.g. the toric code in $2 + 1$ dimensions \cite{Kitaev:1997wr}. In the electric basis this state becomes a sum over \emph{all} representations on all living links, i.e.
\bel{\label{topo}
  \qvec{\trm{topo}} \propto {\prod_\ell}^\star \left(\sum_{r_\ell} d_{r_\ell}  \qvec{r_\ell}^\star_\ell\right).
}
Here $\prod_\ell^\star$ denotes the product over all living links. This is a sum over infinitely many states, and this is reflected in the logarithmic divergence of the entanglement entropy due to the presence of infinitely many superselection sectors with equal probability. This divergence goes back to the bad behavior of the norm, $\qprod{\trm{topo}}{\trm{topo}} = \prod_\ell^\star \delta(0)$, as expected for a state where each link has a definite position. This issue appears in the na\"ive $g = 0$ regime of any gauge theory with continuous gauge group, and as we will now show, it is an artifact of carelessly taking the weak coupling limit.

We must deal with this infinity in order to meaningfully talk about the weak coupling regime. To this end, we regulate the Lie manifold of $G$ with a short-distance cutoff $\eps$. This way the allowed values of $U_\ell$ are of the form $e^{i \eps n^a_\ell T^a}$ with $n^a_\ell \in \Z$. (This is a generalization of regulating the $U(1)$ gauge group with $\Z_\kappa$ and taking $\eps = 1/\kappa$.) In principle, this regularization should have further $\eps^2$ and higher terms in the exponent in order to properly mimic the curvature of the Lie manifold. These corrections will not be important for our purposes, as we will focus on excitations with $|n^a_\ell| \ll 1/\eps$; these will be the ones relevant for describing the physics at very small coupling where the field configurations tend to be very close to $U_\ell = \1$.\footnote{This approximation will be reliable at large $N$, when nonperturbative effects due to the nontrivial topology of the gauge group can be ignored. Conversely, at $N \sim 1$ one should be careful; for instance, for $d = 2$ and $G = U(1)$, monopole effects have a crucial effect on the physics via the Polyakov mechanism \cite{Polyakov:1975}.} The effective Hamiltonian acting on these basis states is found by expanding \eqref{H} in $\eps$:
\bel{\label{def Heff}
  H\_{eff} = - \frac{g^2}{\eps^2} \sum_{\ell,\, a} \left(\frac{\Delta}{\Delta n^a_\ell}\right)^2 + \frac {\eps^2}{g^2} \sum_{p,\, a} \left(\sum_{\ell \in p} n_\ell^a\right)^2.
}
Here we use $\Delta/\Delta n$ to denote the discrete difference operator. The sums over links go over both living and dead links; on each dead link we set $n_\ell^a = 0$ and express $\Delta/\Delta n_\ell^a$ through difference operators on living links. In this limit the colors decouple and we can write $H\_{eff} = \sum_a H^a\_{eff}$. This is the first indication that in the weak coupling regime the theory in a sense behaves as $\trm{dim}(G)$ decoupled photons, which is of course the setup familiar from e.g.~weakly coupled QCD. Because of this we might expect the entanglement entropy to scale with $\trm{dim}(G) \sim N^2$.

The crucial observation that allows us to understand the behavior of $H\_{eff}$ is that there exist two very different extremal regimes, $\eps \gg g$ and $\eps \ll g$. The regime we are in depends on how $\eps$ and $g$ scale as they are both taken to zero. When $\eps \ll g$, the $\eps^2/g^2$ term in the ``magnetic'' part of $H\_{eff}$ is very small, and nearby configurations $n_\ell^a$ and $n_\ell^a + 1$ have infinitesimal energy differences. In other words, we may replace the states $\{\qvec{n^a_\ell}\}$ at fixed $a$ and $\ell$ with a continuous set $\{\qvec{X^a_\ell}\}$ with $X^a_\ell \in (-1/g, 1/g) \approx \R$, so $H\_{eff}$ describes $\dim(G)$ decoupled systems of harmonic oscillators (noncompact photons), each with Hamiltonian $H\_{eff}^a = - \sum_\ell \big(\pder{}{X_\ell^a}\big)^2 + \sum_p (\sum_\ell X_\ell^a)^2$. This is the Coulomb phase, and we will denote the corresponding ground state by $\qvec{\trm{Coulomb}}$. However, when $\eps \gg g$, we cannot so carelessly rescale the fields and get a nice description in terms of continuous oscillators. At finite $\eps/g$, $H\_{eff}$ describes a collection of particles moving in a quadratic potential on a one-dimensional lattice, with $\eps/g$ being the spacing between the sites; as this spacing is taken to infinity, the oscillators all freeze into $n_\ell^a = 0$. This frozen-out configuration is precisely the topological state $\qvec{\trm{topo}}$.\footnote{This regime may alternatively be called a Higgs phase because $\qvec{\trm{topo}}$ breaks the global $G$-symmetry of the physical states by picking $U = \1$. (This symmetry is explicitly broken by the boundary conditions at the edges of the lattice.) Calling this state topological is justified because the equal superposition of loops in $\qvec{\trm{topo}}$ makes the state invariant under arbitrary loop group transformations, which may in turn be viewed as diffeomorphisms on the underlying space.}  The limit $\eps \gg g$ is exactly the regime in which our earlier $g = 0$ discussion was applicable. At finite $\eps/g$ the intermediate ground state wavefunctions can be obtained in terms of Mathieu functions \cite{Chalbaud:1986}.

In order to access the weak coupling regime of a theory with a continuous gauge group, we must take $\eps$ and $g$ small while keeping $\eps \ll g$, and hence we \emph{cannot} na\"ively set $g = 0$ from the outset. If we are working with a discrete gauge group with $\kappa \gg 1$ elements, however, we are at liberty to take the coupling to zero with any ratio $\eps/g \sim 1/\kappa g$; depending on this ratio the ground state interpolates between a topological state and the ground state of weakly coupled noncompact photons. For $\Z_\kappa$ gauge theory this crossover (or transition, depending on $d$) between the Higgs and Coulomb regimes is a classic result \cite{Ukawa:1979yv}.

\section{Abelian gauge theory in $d = 2$} \label{app maxwell}

In this Appendix we review the salient properties of Abelian gauge theories in $d = 2$. This section lies somewhat outside the main line of development of the paper, but we include it for completeness of presentation.

\subsection{Lessons from the continuum}

For most of this section we will focus on the $U(1)$ theory in $d = 2$ spatial dimensions. Let us start from some continuum considerations; a good review is \cite{Dunne:1998qy}. A general renormalizable Lagrangian for the compact $U(1)$ theory has a Maxwell and a Chern-Simons (CS) term,
\bel{\label{def L}
  \L = \frac1{e^2} F_{\mu\nu} F^{\mu\nu} + \frac{i\kappa}{4\pi} \epsilon_{\mu\nu\lambda} A^\mu \del^\nu A^\lambda,\quad A_\mu \sim A_\mu + \frac{2\pi}a,
}
where $\kappa \in \Z$ is the CS level and $a$ is the lattice spacing used to regulate the theory in the UV. In the continuum description we have $A_\mu \in \R$, but the compactness is still recorded by the fact that the physics must be invariant under $A_\mu \mapsto A_\mu + \del_\mu \Lambda$ for $\Lambda \in [0, 2\pi)$. The single propagating degree of freedom in this theory has mass $\kappa e^2$. At distances greater than $\xi = 1/\kappa e^2$ the theory behaves like pure CS at level $\kappa$, and below $\xi$ the theory behaves like the compact Maxwell theory, whose propagators get nonperturbatively screened to zero at lengths $l\_{conf} \sim \exp \frac1{ae^2}$ \cite{Polyakov:1975}. (Typically one has $\xi \ll l\_{conf}$, so this effect is invisible; see \cite{Affleck:1989qf} for a more thorough discussion of monopole screening in the presence of CS terms.) A related theory is the noncompact CS-Maxwell, which has the same Lagrangian $\L$ as \eqref{def L}, except that $\kappa$, $A_\mu$, and $\Lambda$ all take values in $\R$. Physically, the only difference compared to the compact case is that now there is no nonperturbative screening of Maxwell propagators, so at distances below $\xi$ the theory genuinely looks like just a free (noncompact) photon.

This ubiquitous UV/IR structure has interesting consequences in light of the $F$-theorem, which states that the renormalized entanglement entropy $F(r) \equiv r S'(r) - S(r)$ of a circle of radius $r$ is a monotonically decreasing function of $r$. Let $F_{\kappa,\, e}(r)$ be this $F$-function for the CS-Maxwell theory \eqref{def L}, compact or not. The continuum properties discussed above imply that
\bel{\label{eq F}
  F_{\kappa,\, e}(r) = \left\{
                         \begin{array}{ll}
                         F_{0,\, e}(r) , & r \lesssim \xi; \\
                         F_{\kappa,\, \infty}(r) , & r \gtrsim \xi.
                         \end{array}
                       \right.
}
For pure CS it is known that $F_{\kappa,\,\infty}(r) = \frac12 \log \kappa$, so by monotonicity we have $F_{0,\, e}(r) > \frac12 \log \kappa$ for any $\kappa$ and any $r \leq \xi$. This way one can justify finding a logarithmic divergence in the entanglement entropy of small regions in the pure Maxwell theory in the continuum and at any coupling. However, this conclusion is at odds with the entanglement entropy one could calculate for a Maxwell theory on a lattice.

A simple way out is to notice that the CS term is chiral while the Maxwell one is not, and hence the RG flow of pure Maxwell theory will never generate a CS term, and so the above analysis does not say anything about pure Maxwell. However, we can run the above argument for two uncoupled CS-Maxwell theories with opposite levels, or (almost equivalently) for two Maxwell theories coupled with a BF term, and we still have the same conundrum (and no parity arguments to save us).

This tension is resolved in the following way. While the reviewed properties of continuum CS-Maxwell theories are all correct, they are not the whole story. The information missing in \eqref{eq F} is that $F_{\kappa,\, e}(r) = F_{0,\, e}(r)$ holds \emph{at best} only at $a \lesssim r \lesssim \xi$. An easy way to see this is to keep $e$ fixed and increase $\kappa$ until $\xi = 1/\kappa e^2 \sim a$; for any $\kappa > 1/ae^2$ the presence of the CS level will be felt at all length scales and the theory will nowhere behave like pure Maxwell. Conversely, since CS-Maxwell is a massive theory, it must be defined with a UV cutoff and this UV completion will always know about the CS level $\kappa$. In fact, at each $\kappa$ there exists a distinct UV theory at the lattice scale; the free Maxwell theory is not a UV fixed point that controls the flow down to all CS-Maxwell theories, and no CS term will ever be generated by flowing from the pure Maxwell theory. Thus the above $F$-theorem argument only ensures that far enough from both the deep UV and the deep IR --- where the CS-Maxwell theory looks like a pure Maxwell theory --- a term of the form $\log \kappa = -\log(e^2 r)$ appears in $F(r)$. In the deep IR, at $e^2 r \gg 1$, this term becomes $\log \kappa$ (or zero, if there is no CS term), and in the deep UV, at $r \sim a$, the log is replaced by a quantity proportional to $-\log g^2$, the logarithm of the bare coupling. In the main body of the paper we explicitly show that this is a correct prediction and that the entanglement entropy in the continuum Maxwell theory indeed contains the expected, UV-independent logarithmic term.

\subsection{Lessons from the lattice}

We can support the above points by explicitly constructing a lattice theory whose continuum behavior is described by a CS-Maxwell theory at given $\kappa$ and $e$. This construction is standard in condensed matter lore, where it is known that the $U(1) \times U(1)$ CS theory at level $\kappa$ with Lagrangian
\bel{\label{eq BF}
  \frac{i\kappa}{4\pi} \epsilon_{\mu\nu\lambda} \left(A_1^\mu \del^\nu A_2^\lambda + A_1^\mu \del^\nu A_2^\lambda \right)
}
describes the topological phase of the $\Z_\kappa$ lattice gauge theory (see e.g.~\cite{Kou:2008, Hansson:2004} for $\kappa = 2$ incarnations of this idea). Extending this idea, it can be shown that $\Z_\kappa$ gauge theory at coupling $g$ and lattice spacing $a$ is a UV completion of such a $U(1) \times U(1)$ CS-Maxwell theory at level $\kappa$ and coupling $e(g, a)$ for both Maxwell fields.\footnote{Technically, this is a BF-Maxwell theory, but the BF term \eqref{eq BF} can be rewritten as two decoupled CS terms in the weak coupling regime, which is where this theory will be useful for describing a long-distance limit of $\Z_\kappa$, so we will ignore these niceties.} We will spell out below how the IR behavior of $\Z_\kappa$ coincides with the CS-Maxwell theory, but we may already notice that the topological state $\qvec{\trm{topo}}$ of this lattice gauge theory has renormalized entanglement entropy $F(r) = \log \kappa$, which is precisely (and reassuringly) the $F$-function of two CS theories at level $\kappa$.

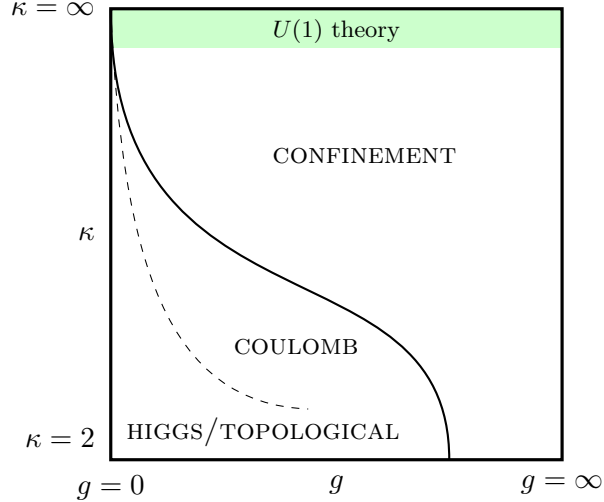
\begin{figure}[tb!]
\begin{center}

\begin{tikzpicture}[scale = 1.5]
  \fill[fill = green!20] (-2, 1.65) rectangle (2, 2);
  \draw[very thick] (-2, -2) rectangle (2, 2);
  \draw[thick] (1, -2) .. controls (1, 0) and  (-2, -1) .. (-2, 2);
  \draw[dashed] (-2, 2) .. controls (-1.9, -0.25) and (-1.5, -1.5).. (-0.25, -1.55);

  \draw (0.25, 0.85) node[anchor = north] {\textsc{confinement}};
  \draw (-1.95, -1.75) node[anchor = west] {\textsc{higgs/topological}};
  \draw (-1, -1) node[anchor = west] {\textsc{coulomb}};

  \draw (0, -2.05) node[anchor = north] {$g$};
  \draw (-2, -2.05) node[anchor = north] {$g = 0$};
  \draw (2, -2.05) node[anchor = north] {$g = \infty$};
  \draw (-2.05, 0) node[anchor = east] {$\kappa$};
  \draw (-2.05, -1.8) node[anchor = east] {$\kappa = 2$};
  \draw (-2.05, 2) node[anchor = east] {$\kappa = \infty$};
  \draw (0, 2) node[anchor = north] {\footnotesize $U(1)$ theory};
\end{tikzpicture}

\end{center}
\caption{\small An impressionistic depiction of the phase structure of $\Z_\kappa$ gauge theory in $d = 2$. The level $\kappa$ is an integer and is represented as continuous just for convenience. The thick line roughly connects critical couplings at which deconfinement happens (the transition is second order for $\kappa = 2$ but may be first order for other levels). The dashed line follows $\kappa = 1/g$, roughly indicating where the weakly coupled ground state crosses over from a noncompact photon to the topological state. The shaded region, given by $\kappa \gg 1$ and $g \gg 1/\kappa$, is where the theory looks like a compact Maxwell lattice theory. At $1 \gg g \gg 1/\kappa$ (the leftmost part of the shaded region) the theory confines at very large distances due to the Polyakov mechanism, and at distances below this confinement scale the theory behaves as noncompact Maxwell, just as in the Coulomb phase.}
\label{fig phases}
\end{figure}

The phase diagram of the $\Z_\kappa$ theory can be worked out by the machinery developed in the previous two sections. We set the lattice spacing to be $a = 1$. At any $\kappa$ the theory confines at large enough $g$. At $1\gg g \gg \eps = 1/\kappa$, the theory at short distances appears to be in the Coulomb phase. However, in this limit $\Z_\kappa$ begins to look like $U(1)$, so the Polyakov mechanism gives the photons a very small mass gap and screens them at very long distances, meaning that the theory is actually confining. As $\eps/g$ is dialed away from zero and towards infinity, the Coulomb phase crosses over to the topological phase. At high enough $\kappa$, the would-be-Coulomb-but-actually-confined phase undergoes a deconfinement transition, has a very short transient behavior, and settles into the topological phase. This behavior is shown on Fig.~\ref{fig phases}.

These phases precisely translate into the previously described regimes of continuum CS-Maxwell. As we have repeatedly emphasized, in the $\eps \ll g$ region the gauge group is effectively $U(1)$ and the continuum theory is compact Maxwell, which is always in the confined phase. When the coupling $g$ is weak and $\eps/g \sim 1$, the oscillations about the ground state start becoming suppressed because the photons are restricted to take values on a discrete grid of spacing $\eps/g$; this makes the photon in the Coulomb phase massive, and this corresponds to giving it mass $\kappa e^2$ in the continuum, which is the hallmark of the CS-Maxwell theory. (It would be interesting to work out exactly how $\eps/g$ translates to $\kappa e^2$.) Finally, as the topological phase is reached at $\eps \gg g$, all the fluctuations become infinitely gapped and we are left with the topological state described by the pure CS theory in the continuum.

The Maxwell theory on a lattice, being obtained by sending $\kappa \rar \infty$ before taking any other limits of the $\Z_\kappa$ theory, is thus not described by an effective CS theory even in the deep IR. In particular, the compact Maxwell theory on a lattice will always be trivial in the IR due to the Polyakov mechanism, and the noncompact Maxwell theory on a lattice will be scale-invariant and will not RG flow.

Even though the pure Maxwell theory on a lattice does not have a topological phase, coupling it to other theories can allow it to flow to something topological in the IR. In particular, coupling Maxwell theory to scalars of charge $\kappa \geq 2$ can Higgs it and lead to a $\Z_\kappa$ gauge theory \cite{Fradkin:1978dv}. In the continuum language, this means that scalar QED$_3$ can be written as a purely topological theory in the IR, which is a fact often used in studies of the fractional quantum Hall effect \cite{Zee:1996fe}. Conversely, coupling a pure Maxwell theory to a massive fermion will lead to just a $U(1)$ CS theory, and indeed coupling any gauge theory to fermions is the right way to access the chiral CS regime, even for non-Abelian groups \cite{Giombi:2013yva}. It would be fascinating to extend the work in this paper to gauge-matter theories and to verify that the $F$-theorem holds for them too.

\section{Maxwell-scalar duality in $d = 2$} \label{app dual}

In the Hamiltonian formalism on a $d = 2$ lattice, the Maxwell-scalar duality is the operator map
\algnl{\label{dual 1}
  g J_{\ell(p_1, p_2)} &\equiv \phi_{p_1} - \phi_{p_2},\\ \label{dual 2}
  \sum_{\ell \in p} A_\ell &\equiv g \pi_p,
}
where $\ell(p_1, p_2)$ is the link between adjacent plaquettes $p_1$ and $p_2$. The conjugate operators $\phi_p$ and $\pi_p$ describe a scalar theory on the dual lattice. At small $g$, the Hamiltonian is that of a free massless field, $\frac12 \sum_p \pi_p^2 + \frac12 \sum_{\avg{p_1, p_2}} (\phi_{p_1} - \phi_{p_2})^2$. Eigenvalues of $J_\ell$ are integers, and hence $\phi_p$ has eigenvalues in $g\Z$, so at small gauge coupling the dual scalar takes values in a continuous set. The definition above specifies $\phi_p$ up to a global shift, $\phi_p \mapsto \phi_p + g$. This is crucial: it means that the dual scalar theory is \emph{not} a simple free field, but rather a ``truncated scalar'' (a noncompact photon or a Nambu-Goldstone boson), i.e.~a scalar field in which configurations related by global shifts by $g$ have been identified. This ``gauging'' of the shift symmetry will lead to the promised appearance of log terms in the entanglement entropy.

In the continuum limit, the mapping \eqref{dual 1}, \eqref{dual 2} must first be amended by replacing $g$ with $g/a^{1/2}$. Taking $g \rar 0$ and $a \rar 0$ with $e^2 = g^2/a$ and employing the appropriate continuum variables, the duality takes the well-known form
\algnl{
  J_i(\b x) &\equiv \frac1e \del_i\phi(\b x),\\
  \epsilon^{ij} \del_i A_j(\b x) &\equiv e \pi(\b x),
}
or $\frac12 \epsilon_{\mu\nu\lambda} F^{\mu\nu} \equiv e \del_\lambda \phi$ for short. The electric operator $J_i(\b x)$ has real eigenvalues so $\phi(\b x)$ is also real, but the shift identification becomes $\phi(\b x) \equiv \phi(\b x) + e$, meaning that the dual description of Maxwell theory is the spontaneous symmetry breaking phase of a compact scalar with radius $e$.

When the continuum coupling is small (i.e.~at length scales $r$ such that $e^2 r \ll 1$) the shift identification reduces to the gauging of infinitesimal shifts $\phi(\b x) \mapsto \phi(\b x) + \eps$. This is equivalent to removing the zero mode from the theory. Qualitatively, the entanglement entropy at such small scales will reflect the loss of this one mode \cite{Kulchytskyy:2015}; the softest (nearly constant) mode that can live in the entangling region will be forced to couple to the softest mode allowed in the exterior in order to put the overall zero mode into its ground state, a state it has to stay in because it plays no part at small coupling. As the entangling region is increased, its softest mode will be allowed more and more leeway, leading to the presence of a monotonically increasing $\log(e^2 r)$ term in the entanglement entropy, as we will show explicitly below. As we work our way to higher length scales, only finite shifts will be gauged away and the logarithmic term will cross over to an $O(1)$ constant by the time we reach the $e^2 r \gg 1$ regime, where there are effectively no traces of the gauging left and the entanglement entropy corresponds to that of the noncompact, ordinary scalar. By further increasing $e^2 r$ we will eventually hit the confinement scale $r = l\_{conf}$ at which point the dual scalar becomes massive and the gauge fields become confined, and the entanglement entropy becomes zero.

We close this section by remarking that the dual photon $\phi$ is \emph{not} the same field as the gauge-fixed vector potential, even though both can be written as a single real degree of freedom. Given a plaquette with living links $\ell_1$ and $\ell_2$, the two descriptions are related by eq.~\eqref{dual 2},
\bel{
  A_{\ell_1} - A_{\ell_2} = - ig \pder{}{\phi_p}.
}
In other words, axial gauge is related to the dual photon by a conjugate transformation that exchanges the position and momentum operators.

\end{document}